\documentclass[pre,superscriptaddress,twocolumn,amsfonts,amssymb,amsmath,floatfix,showpacs]{revtex4}

\usepackage{graphicx}
\usepackage{bm}

\begin{document}
\newlength{\GraphicsWidth}
\setlength{\GraphicsWidth}{8cm}
\newcommand{\Label}[1]{\label{#1}}                  
 \newcommand{\Bibitem}[1]{\bibitem{#1}\{#1\}}     

\newcommand\be{\begin{equation}}
\newcommand\ee{\end{equation}}
\newcommand\ba{\begin{eqnarray}}
\newcommand\ea{\end{eqnarray}}
\newcommand{\nn}{\nonumber\\}
\newcommand{\av}[1]{\langle #1\rangle}  
\newcommand{\dav}[1]{\langle\langle #1\rangle\rangle}
\newcommand{\inprod}[2]{\langle #1 | #2 \rangle}
\newcommand{\half}{\textstyle{\frac{1}{2}}}
\newcommand{\third}{\textstyle{\frac{1}{3}}}
\newcommand{\fourth}{\textstyle{\frac{1}{4}}}
\newcommand{\al}{{\alpha}}
\newcommand{\bhat}[1]{\hat{{\bf #1}}}
\newcommand{\tbkmin}{\tilde{{\bf k}}_-}
\newcommand{\tbkplus}{\tilde{{\bf k}}_+}
\newcommand{\tkmin}{\tilde{{k}}_-}
\newcommand{\tkplus}{\tilde{{k}}_+}
\newcommand{\tf}{\tilde{f}}
\newcommand{\tI}{\tilde{I}}
\newcommand{\hbv}{\widehat{{\bf v}}}
\newcommand{\hbk}{\widehat{{\bf k}}}
\newcommand{\bkplus}{{{\bf k}}_+}
\newcommand{\bkmin}{{{\bf k}}_-}
\newcommand{\bk}{{\bf k}}
\newcommand{\bq}{{\bf q}}
\newcommand{\bg}{{\bf g}}
\newcommand{\bgperp}{{{\bg}_\bot}}
\newcommand{\bn}{{\bf n}}
\newcommand{\bG}{{\bf G}}
\newcommand{\rg}{{\rm g}}
\newcommand{\gperp}{{\rg_\bot}}
\newcommand{\gpar}{{{\rm g}_\parallel}}
\newcommand{\hgpar}{{\hat{\rg}_\parallel}}
\newcommand{\bv}{{\bf v}}
\newcommand{\bc}{{\bf c}}
\newcommand{\bw}{{\bf w}}
\newcommand{\br}{{\bf r}}
\newcommand{\eps}{{\varepsilon}}
\newcommand{\Ref}[1]{(\ref{#1})}
\newcommand{\bull}{{$\bullet$}}
\newcommand{\XXXX}{\bf XXXX }

\newcommand{\combin}[2]{\left( \begin{array}{c} #1 \\
                        #2 \end{array} \right)}

\title{
Boltzmann equation for dissipative gases in homogeneous states
with nonlinear friction}
\author{E. Trizac}
\affiliation{Univ Paris-Sud, 91405 Orsay, France}
\affiliation{LPTMS (UMR CNRS 8626), 91405 Orsay, France}
\author{A. Barrat}
\affiliation{Univ Paris-Sud, 91405 Orsay, France}
\affiliation{LPT (CNRS, UMR 8627), 91405 Orsay, France}
\author{M.H. Ernst}
\affiliation{Instituut voor Theoretische Fysica, Universiteit Utrecht,
Postbus 80.195, 3508 TD Utrecht (The Netherlands)}

\begin{abstract}
  Combining analytical and numerical methods, we study
  within the framework of the homogeneous non-linear Boltzmann equation,
  a broad class of models
  relevant for the dynamics of dissipative fluids, including granular gases.
  We use the new method presented in a previous paper [J. Stat. Phys.
  {\bf 124}, 549 (2006)] and extend our results to a different heating
  mechanism, namely a deterministic non-linear friction force. We derive
  analytically the high energy tail of the velocity distribution and compare the
  theoretical predictions with high precision numerical simulations. Stretched
  exponential forms are obtained when the non-equilibrium steady state is
  stable. We derive sub-leading corrections and emphasize their relevance. In
  marginal stability cases, power-law behaviors arise, with exponents
  obtained as the roots of transcendental equations. We also consider some
  simple BGK (Bhatnagar, Gross, Krook) models, driven by similar heating
  devices, to test the robustness of our predictions.
\end{abstract}


\maketitle


\section{Introduction}

Granular materials represent one of the most well-known paradigms for open
dissipative systems; their ubiquitous character in natural phenomena or
industrial processes, the possibility to study them at both very applied and
very fundamental levels have prompted many efforts to understand their
properties \cite{Jaeger,PoschelBrill,barrat05,poschel01,poschel03}, but they
remain challenging from many points of view. Thermodynamic-like descriptions
remain in particular elusive for these intrinsically far from equilibrium
systems, because energy is continuously lost through internal dissipation,
and has to be compensated by non-thermal sources.

Dilute granular gases present a particularly interesting framework which can
be studied by model experiments, numerical simulations, hydrodynamics,
kinetic theory or more phenomenological approaches.
The investigation of the velocity distribution of the
particles reveals a rich phenomenology, with strong deviations from the
equilibrium Maxwell-Boltzmann behavior, and its description is still an object
of debates. From an experimental point of view, the situation may appear
confusing at first sight. Although the measured velocity distribution $F({\bf
  v})$ generically deviates from the Maxwellian, its
functional form depends on material property and on the forcing
mechanism used to compensate for collisional loss of energy
\cite{Exp,Prevost,EPJE,Aranson}. A similar picture emerges from
numerical simulations and analytical studies
\cite{Aranson,Gold,Brey,puglisi,Rome,Fuchs,stretch-tails,MS,Cafiero,Swinney,Santos,Maxwell,KBN02,EB-JSP,EB-PRE,Math,BBRTvW,Pias,ben-Avb,EPJE,Zip,vanZon,Haifa,EBN-Machta,JSP-1,EPL},
where in addition to the more common stretched exponential
behavior, power-law distributions have also been reported
\cite{Rome,Maxwell,KBN02,EB-JSP,EBN-Machta}.

Since the universality of the Maxwell-Boltzmann distribution for
equilibrium gases seems to have no analog in steady states of dissipative
gases, it is necessary to identify the generic trends for the velocity
distribution and study their properties in a unified and simple framework.
Our purpose is to develop further an existing
quantitative cartoon to unveil the different
effects at work, that lead to the wide range of behaviors alluded to
above, when certain key parameters are changed. To this aim, we focus on the
framework of the generalized inelastic Boltzmann equation, which describes
dilute homogeneous dissipative gases. The study of homogeneous systems is
not only a useful starting point, it is also relevant to experiments with
bulk driving \cite{Prevost,Aranson}. Spatial heterogeneity (and thus
gravity, hydrodynamic instabilities, shock formation and clustering
\cite{PoschelBrill}) will henceforth be discarded.

The most characteristic features of the velocity distribution in
dissipative systems -- whether observed in time dependent scaling
states or in non-equilibrium steady states -- is overpopulated
high energy tails. Generically these tails are stretched
exponentials \cite{stretch-tails,MS,EB-PRE}. The more spectacular
power law tails \cite{Rome,Maxwell,KBN02,EB-JSP,EBN-Machta} are
exceptional. They were only found in systems of Maxwell molecules,
and in the unusual device of Ref.\cite{EBN-Machta}. In general
these power law tails do not evolve naturally, but require careful
fine-tuning of the physical parameters, both in the interactions,
{\it and } in the driving mechanisms \cite{JSP-1,EPL}.

In the present paper, specific features and properties of $F(v)$
will be derived and analyzed, when energy is injected by a
negative friction thermostat, parametrized by an exponent
$\theta$.
An important new result of general importance \cite{EPL,JSP-1}
is also the direct and simple relation between the
parameters controlling the stability of the energy
balance equation, i.e. the balance between collisional
dissipation and energy injection on the one hand, and
on the other hand the occurrence of new power law tails
$\sim 1/v^{a(\theta)}$
in the velocity distribution of the non equilibrium steady state, 
which appear
at the margin of stability in this equation.
In Ref. \cite{EPL} a short summary of
our results was presented. In this longer paper we show how all
these results, and in particular these new power law tails,
have been obtained
with the help of a new method developed in Ref.\cite{EPL,JSP-1},
which is applicable to different types of forcing mechanisms, and
to a large class of interaction models, including inelastic hard
spheres and Maxwell molecules. It seems virtually impossible to
obtain the present results by using the old methods developed for
Maxwell molecules \cite{Rome,Maxwell,EB-JSP}.

The current paper hence constitutes a sequel to our first study
\cite{JSP-1}, in which we have developed a method for analyzing
the deviations of $F(\bf v)$ from Gaussian behavior. While we have
focused in \cite{JSP-1} on granular gases for which energy is
injected by random forces, we consider here a different driving
mechanism. We show that the large velocity tail of $F(\bv)$ is
characterized by a stretched exponential $\exp[-v^b]$ with
an exponent $b$, that governs the stability of the non-equilibrium
steady state. When this state is a stable fixed point of the
dynamics, the exponent satisfies $b>0$. For the
exceptional cases of marginal stability, where $b$ vanishes, $F$
is of power law type, and the corresponding exponents will be
calculated.

The paper is organized as follows: we recall in section \ref{sec2} the
Boltzmann equation for inelastic soft spheres, together with the criterion
of stability of non-equilibrium steady states (NESS). We also briefly
recall in \ref{subsec2b} the method introduced in our previous paper
\cite{JSP-1}. Sections \ref{sec3} and \ref{sec4} give the results for the
large velocity tail of the velocity distribution in the case of energy
injection by non-linear negative friction, for both stable and marginally
stable NESS. We turn in section \ref{sec6} to
a simple linear model which mimics the basics of the inelastic Boltzmann
equation. While this BGK model is amenable to analytical treatment, we
show that it does not reproduce the rich behavior of the non-linear
Boltzmann equation, in particular it fails for hard interactions.
Section \ref{sec:concl} finally contains our conclusions.

\section{Inelastic Boltzmann equation}
\label{sec2}

We consider the Boltzmann equation for a gas of inelastic soft spheres
\cite{JSP-1}, which represents one of the simplest models for rapid
granular flows. The dynamics is described as a succession of uncorrelated
inelastic binary collisions, modelled by soft spheres with a collision
frequency and a coefficient of normal restitution $\alpha$, where $
0<\alpha <1$ \cite{PoschelBrill}. The  collision law $(\bv_1,\bv_2)\to
(\bv'_1,\bf v'_2)$ reads:
\be
\bv'_{1} = \bv_{1} -  p({\bf{g}}\!\cdot \! {\bf{n}}) { \bf{n}}, \quad
{\bv'_{2}} = {\bv_{2}} +  {p}({\bf{g}}\!\cdot \! {\bf{n}})  \bf{n}
\label{v-coll}
\ee
where $\bf{g} \equiv \bv_1-\bv_2$, $ p= 1-q =\frac{1}{2}(1+\alpha)$ and
$\bf{n}$ is a unit vector parallel to the impact direction connecting
particles 1 and 2. Inelastic collisions conserve mass and momentum, and
dissipate kinetic energy at a rate $\propto {1\over 2}(1-\alpha^2)= 2pq$
(the elastic case corresponds to $\alpha=1$). We consider a general
collision frequency, $g \varsigma (g,\vartheta) \sim g^\nu
|\widehat{\boldsymbol g}\cdot \bf{n}|^\sigma$. Here $\varsigma
(g,\vartheta) $ is the differential scattering cross section with
$\vartheta = \cos^{-1}(\widehat{\boldsymbol g}\cdot \bf{n} )$, $\nu$
describes its energy dependence, and $\sigma $ its angular dependence.
The exponent $\sigma \neq 1$ describes a
distribution of impact parameters biased towards grazing ($ \sigma < 1$)
or head-on ($ \sigma> 1$) collisions. 
For mathematical
convenience models with $ \sigma =\nu $ have also been considered
\cite{KBN02,EB-JSP,JSP-1}. The symbol
$\mathbf{\widehat{a}} = {\bf a} /a $ denotes a unit vector, parallel to
$\bf{a}$. For elastic particles interacting via a soft sphere potential
$U(r)\propto r^{-a}$, one has $\nu = 1-2(d-1)/a$ \cite{Phys-Rep-ME}, where
$d$ is the space dimension. The exponents $\nu= \sigma=1$ correspond to
standard hard-sphere behavior $(a\to \infty)$, while $\nu=0$ corresponds
to Maxwell molecules ($a=2(d-1)$). Here $\nu$ and $\sigma$ will be free
exponents, that parametrize the material properties together with
the inelasticity parameter $\alpha$.

We now give a simple representation of the nonlinear Boltzmann collision
operator, which is convenient to
study the spectral properties of the linearized collision operator. The
time dependent  distribution $F(\bv,t)$  in spatially homogeneous systems
obeys the nonlinear Boltzmann equation \cite{EPL}:
\begin{eqnarray} \label{BE}
&\partial_t F(\bv,t) + {\cal F}F = I(v|F) \equiv \int_\bn \int d\bv_1
d\bv_2  g^\nu |\widehat{\bf{g}}\cdot\bf{n}|^\sigma \times & \nn
 & F(\bv_1,t) F(\bv_2,t) [\delta(\bv-\bv'_1)-
\delta(\bv-\bv_1)]&
\end{eqnarray}
where the collision operator $I(v|F)$ has the usual gain-loss structure.
For anisotropic $F(\bv, t)$ the angular integral, $\int_\bn (\cdots) =
\int^{(-)} d \bn (\cdots) /\int^{(-)} d \bn $ denotes an average over the
pre-collision hemisphere, $\mathbf{g}\cdot{\bf{n}}\leq0$. For isotropic
distributions, as considered here, the integrals over pre- and
post-collision hemisphere are the same, and $\int_\bn (\cdots) = \int d
\bn (\cdots) /\Omega_d $ can be extended over the complete solid angle,
i.e.  where $\Omega_d = 2 \pi^{d/2} /\Gamma (d/2) $ is the surface area of
a $d$-dimensional unit sphere, and $\Gamma(x)$ is the Gamma function.

The forcing term ${\cal F}F$ represents the energy supply, working against
inelastic dissipation. It may lead to a NESS (Non-Equilibrium Steady
State). Absence of forcing ($\mathcal{F} =0$) describes free cooling,
where the energy is decreasing in time.  A heating device, considered
frequently, consists in a random force acting on the particles in between
collisions: the corresponding {\it stochastic} White Noise (WN), widely
used in analytical and numerical studies
\cite{stretch-tails,Pago,MS,Swinney,BBRTvW,vanZon}, is described by adding a
diffusion term $-D {\bf\partial}_\bv^2 F$ to the Boltzmann equation.  Our
previous paper \cite{JSP-1} has focused on this case. An interesting
alternative is given by {\it deterministic} nonlinear Negative Friction
(NF).  In general, the forcing term reads
\be \label{forcing} {\cal F}F = ({\cal F}_{\rm NF} + {\cal F}_{\rm WN})F
=\gamma \partial_{\bv}\cdot(\bv v^{\theta-1} F) \,-\, D \,
\partial_{\bv}^2 F
\ee
with ($\gamma=0$, $D\neq0$) for WN and
($\gamma >0$, $D=0$) for NF, and $\theta \geq 0$. The
value $\theta=1$ corresponding to the Gaussian
thermostat allows us to study the long time scaling regime of an unforced
system (so-called free cooling) (see section
\ref{subsec2a} and \cite{MS}). The special case $\theta=0$ models
gliding friction. In general $\theta$ is a continuous exponent selectively
controlling the energy injection mechanism. In summary, the 'phase space' to
explore is thus given by the parameters ($\nu,\sigma,\theta,\alpha$) for NF
driving, which is the main focus of this paper.

The explicit form of $ I(v|F)$ above is convenient for calculating
the rate of change of averages, $\int {d\bv} \psi(v) F(v,t)\equiv \langle
\psi|F \rangle_t$, i.e.
\ba \label{rate-av}
\partial_t \:\langle \psi(\bv)|F \rangle_t + \langle\psi |{\cal F}F\rangle_t =&
\frac{1}{2}& \int_\bn \int d\bv_1 d\bv_2 \: g^\nu
|\widehat{\bf{g}}\cdot\bf{n}|^\sigma  \times \nn F(\bv_1,t) F(\bv_2,t)
[\psi(\bv'_1) + \psi(\!&\!\bv'_2\!&\!)-\psi(\bv_1)- \psi(\bv_2)]
\ea
Here \Ref{BE} is just a special case of \Ref{rate-av}, where
$\langle\psi(\bv)|F\rangle_t  = F(\bw,t)$ for $\psi (\bv) = \delta (\bv
-\bw)$. The loss rate of energy follows by setting $\av{\psi(\bv)|F}_t =
\av{\frac{1}{2} m v^2|F}_t \equiv \frac{d}{4} m v_0^2(t)$, where $v_0(t)$
is the r.m.s. velocity and $v^2_0$ the granular temperature. We further
use the normalizations, $ \av {(1, \bv, v^2)|F}_t = (1,0,d v_0^2(t)/2)$.

\subsection{Scaling form and stability of steady states}
\label{subsec2a}

To analyze the behavior of the distribution function we assume a rapid
approach to a scaling form,
\be \label{scaling-f} F(v,t) = (1/v_0(t))^d
f(v/v_0(t)) \ee (for related proofs, see \cite{Math}) with normalizations \be
\label{norm-f} \langle (1, c^2)|f\rangle = (1, d/2).
\ee
The inelastic Boltzmann equation \Ref{BE} can then be decoupled into a
time-independent equation for the scaling form $f(c)$
and a time-dependent equation for the r.m.s. velocity $v_0$, or granular
temperature $v^2_0$, which reads
\be \label{v0}
\frac{d}{dt}\left( \frac{d}{2} v^2_0\right) = \av{v^2|I}_t
-\av{v^2|\mathcal{F}F}_t .
\ee
The collisional average $ \av{v^2|I}_t$ follows from \Ref{rate-av} and
\Ref{v-coll} by carrying out the angular average, and by inserting
\Ref{scaling-f} to change to scaling variables. Here $\int_\bn
|\widehat{\bg} \cdot \bn|^{\sigma+2} =\beta_{\sigma+2}$ is given by
\be
\label{cos-av}
\beta_\sigma =
\int_\bn |\widehat{\bg}\cdot \bn|^\sigma
= \textstyle{\Gamma(\frac{\sigma+1}{2}) \Gamma(\frac{d}{2})/
\Gamma(\frac{\sigma+d}{2})\Gamma(\frac{1}{2}) }
\ee
 where $\beta_\sigma$ is well defined for $\sigma > -1$. The forcing terms, $
\av{v^2|\mathcal{F}F}_t$ are calculated along the same lines using
\Ref{forcing} and partial integrations. The results are,
\ba \label{coll-av}
&&\av{v^2|I}_t = - \textstyle{\frac{1}{2}} \lambda_2
\dav{g^{\nu+2}} v_0^{\nu+2}(t)
\nn
&&\av{v^2|\mathcal{F}_{\rm WN}F}_t = -2dD
\nn
&&\av{v^2|\mathcal{F}_{\rm NF}F}_t = -2\gamma
\av{c^{\theta+1}}v_0^{\theta+1}(t),
\ea
where $g =|\bc_1-\bc_2|$. The coefficient $\lambda_2 =
2pq\beta_{\sigma+2}$ can be identified as the eigenvalue of
the linearized Boltzmann collision operator (see 
the general expression for $\lambda_s$ below, Eq. (\Ref{eigen-val}) 
next subsection). Here,
the notation $\dav{k(\bc_1,\bc_2)}$ stands for the average of a function
$k(\bc_1,\bc_2)$ with weight $f(\bc_1) f(\bc_2)$.

Let us first consider forced systems. The two terms in (\ref{v0}) can then
balance each other and lead to a NESS. For the WN-driven case
($\gamma=0$), Eqs.\Ref{v0}-\Ref{coll-av} give
\ba \label{rate-v0-WN}
dv^2_0/dt &=& 4D -\textstyle{\frac{1}{d}}\lambda_2
\dav{g^{\nu+2}} v_0^{\nu +2}(t)
\nn &=& 4D \left[1-\left(\textstyle{\frac{v_0(t)}{v_0(\infty)}}
\right)^{2b}\right]
\ea
where $b_{\rm WN} = 1 +\textstyle{\frac{1}{2}} \nu$, and
$v_0(\infty)$ is defined as the stationary solution of
\Ref{rate-v0-WN}. Similarly we obtain for the NF-case,
\ba \label{rate-v0-NF}
dv_0/dt &=& (\textstyle{\frac{2\gamma}{d}}) \av{c^{\theta+1}} v_0^\theta
(t) -\textstyle{\frac{\lambda_2}{2d}} \dav{g^{\nu+2}} v_0^{\nu +1}(t)
\nn &=&(\textstyle{\frac{2\gamma}{d}}) \av{c^{\theta+1}}v_0^\theta (t)
\left[1-\left(\frac{v_0(t)}{v_0(\infty)}
\right)^{b}\right]
\ea
where $b_{\rm NF} =  \nu +1 -\theta$. The dynamics always admits a
fixed point solution of the equations above. The fixed point
solution $v_0(\infty)$ is stable/attracting for $b>0$,
unstable/repelling for $b<0$, and marginally stable for $b=0$. As
long as $b>0$, the system naturally evolves towards the stable NESS.
Note that $D$ and $\gamma$ in
\Ref{rate-v0-WN} and \Ref{rate-v0-NF} are irrelevant phenomenological
constants, that can be absorbed in energy and time scales.

In the stable NESS $(\dot{v}_0 =0, b>0)$, the corresponding
integral equations for $f(c)$ follow from \Ref{BE} with $\partial_t
F=0$ and \Ref{forcing}, and yield respectively for WN and NF,
\begin{eqnarray} 
&I(c|f) = -\frac{D}{(v_0(\infty))^{2b}} \partial^2_{{\bf c}} f =
-\frac{\lambda_2}{4d}
\langle\langle g^{\nu+2}\rangle\rangle \partial^2_{{\bf c}} f &
\\
&I(c|f) = \frac{\gamma}{(v_0(\infty))^b} \partial_{{\bf c}}\!\cdot
\!(\widehat{{\bf c}} c^\theta f) =  \frac{\lambda_2
\langle\langle g^{\nu+2}\rangle\rangle}{4 \langle c^{\theta
+1}\rangle}{\partial}_{{\bf c}}\!\cdot \! (\widehat{{\bf c}}
c^\theta f)& \label{int}
\end{eqnarray}
Here $D$ and $\gamma$ have been eliminated with the help of the
steady state solutions of \Ref{rate-v0-WN} and \Ref{rate-v0-NF}.

It is also possible to consider the freely evolving state (FC), that does
not lead to a NESS since $\gamma=0$ and $D=0$, but to a scaling solution
$f(c)$. Here the r.m.s. velocity decays according to
\be \label{rate-v0-FC}
\dot{v}_0 =  -\textstyle{\frac{\lambda_2}{2d}} \dav{g^{\nu+2}} v_0^{\nu
+1}(t) .
\ee
Moreover comparison of the FC case with the Gaussian thermostat (NF:
$\theta=1$) shows  that the corresponding integral equations for $f(c)$ are
identical, as observed in \cite{MS}. This occurs because the term
arising from $\partial_t F$ in \Ref{BE} for FC is non-zero (since we do
not have a NESS) and  corresponds exactly with the forcing term in Eq.
\Ref{int} when $\theta=1$ (since $\langle c^2\rangle=d/2$). The Gaussian
thermostat in fact describes a system driven by a {\it linear} (negative)
friction force, ${\bf a} = \gamma \bc$. This corresponds to a linear
rescaling of the velocities, which leaves $f(c)$ invariant. 
While the NF with $theta=1$ and FC states are equivalent at the level of
the scaled velocity distribution function, they differ in the evolution
equation for the temperature. It is therefore not possible
to extend the stability criterion of NF to FC systems.

We also note that \Ref{rate-v0-FC} gives a generalization of Haff's
law, that describes the decay of the energy or granular temperature
for inelastic soft spheres. For instance, for $\nu>0 $ one finds
$v^2_0(t) \sim t^{-2/\nu}$ (for more details, see \cite{JSP-1}).

\subsection{Linearized Boltzmann operator}
\label{subsec2b}

We now recall briefly the method introduced in \cite{JSP-1}, which allows
us to obtain the  behavior of the velocity distributions at large
velocities. Suppose that for large $c = v/v_0$ the velocity distribution
can be separated into two parts, $f(c) = f_0(c) + h(c)$, where $h(c)$ is
the {\it singular} tail part that we want to determine, and $f_0(c)$ the
presumably {\it regular} bulk part. The tail part $h(c)$ may be
exponentially bound $\sim \exp[-\beta c^b] $ with $0<b<2$, or of power law
type. In the bulk part $f_0(c)$ the variable $c$ is effectively restricted
to bulk values in the thermal range $v \lesssim v_0$ or $c \lesssim 1$. As
far as large velocities are concerned, the thermal range of $f$ may be
viewed to zeroth approximation as a Dirac delta function $\delta({\bf
c})$, carrying all the mass of the distribution. In this way, we obtain
an asymptotic expansion of $f(c)$ by considering the ansatz
\be \label{d+h}
f(c)=\delta(\bc) + h(c),
\ee
and linearizing the collision term $I(c|f)$
around the delta function (using the relation $ I(c|\delta) =0$). This
defines the linearized Boltzmann collision operator,
\be
\label{Lambda} I(c|\delta + h) = - \Lambda h(c) + {\cal O}(h^2).
\ee
Note that we restrict the analysis to isotropic functions $h(c)$. The
eigenfunctions of $\Lambda$ decay like powers $c^{-s}$.  Consequently they
are very suitable for describing power law tails, $f(c) \sim c^{-s}$. We
note, however, that the moments $\int d\bc c^s f(c)$ are largely
determined by the {\it regular} bulk part, which can therefore {\it not}
be approximated by $\delta (\bc)$ when calculating moments.

The most important spectral properties are the eigenvalues and
right and left eigenfunctions,
\be \label{eigen-f}
\Lambda c^{-s-d-\nu} = \lambda_sc^{-s-d} \quad;\quad
 \Lambda^\dagger  c^s  = \lambda_s c^{s+\nu}
\ee
with eigenvalues for $s>0$ \cite{JSP-1},
\be \label{eigen-val}
\lambda_s = \beta_\sigma \left\{ 1
- _2\!F_1\,\left(\textstyle{-\frac{s}{2},
\frac{\sigma+1}{2};\frac{\sigma +d}{2}} \,|\,1-q^2\right)\right\} -p^s
\beta_{s+\sigma}.
\ee
Here $_2F_1(a,b;c|z)$ is a hyper-geometric function and $\beta_\sigma$ is
given by (\ref{cos-av}). The value $s=0$ with $\lambda_0=0$ is an isolated
point of the spectrum with corresponding stationary eigenfunctions,
invariant under collisions,
\be \label{zero-eigen-f}
\Lambda \delta(\bc) = 0 \qquad \mbox{and} \qquad \Lambda^\dagger \cdot 1 = 0
\qquad (\lambda_0=0).
\ee
Right and left eigenfunctions are different because $\Lambda$ is
not self-adjoint. There is in fact among the eigenfunctions in
\Ref{eigen-f} another, less trivial, stationary right eigenfunction,
$c^{-s^*-d-\nu}$, where $s^*$ is the root of transcendental
equation $\lambda_s=0$ (see \cite{JSP-1}).

\begin{figure}
\includegraphics[angle=0,width=7cm]{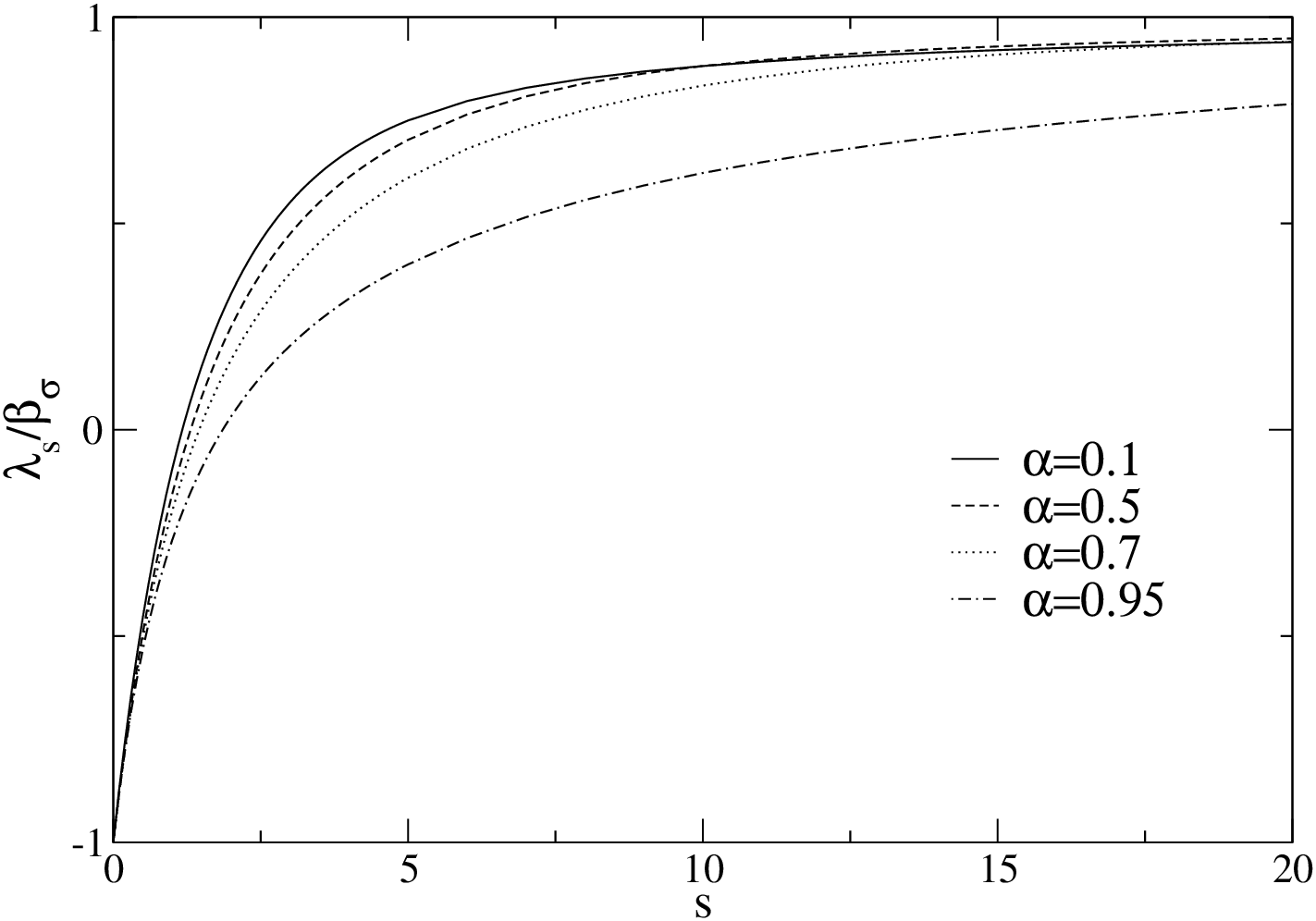}
\caption{
Concave eigenvalue spectrum $\lambda_s(\sigma)$ for $s \geq 0$ of the
collision operator in the inelastic soft sphere models $\{\sigma,\nu\}$,
defined in the text, and shown for various values of the coefficient of
restitution $\alpha$. The ordinate shows $\lambda_s(\sigma)/\beta_\sigma$
for $d=2,\sigma=\nu=1$, which approaches 1 for $s \to \infty$, and -1 for
$s \to 0$. The point $\lambda_0(\sigma)=0$ is an isolated point of the
spectrum.} \label{figlambdas}
\end{figure}

We also note that the eigenvalue $\lambda_s$ is {\em independent} of the
energy exponent $\nu$: it is the same for inelastic Maxwell molecules,
hard spheres, very hard particles, and very weakly interacting particles.
The reason is presumably that the scattering laws are the same in all
models, and equal to those of inelastic hard spheres. Moreover, it depends
strongly  on the inelasticity through $\alpha$, and weakly on the angular
exponent $\sigma$. Fig. \ref{figlambdas} shows that $\lambda_s$ is a
concave function of $s$.

The previous paragraphs refer to the action of the $I(c|f)$ on functions
$f(c)$ of power law type. We shall also need the large-$c$ form,
$I_\infty$, of the collision operator acting on exponentially bound
functions, $ f(c)\sim \exp[-\beta c^b]$, with positive constants $\beta,
b$. The nonlinear operator reduces to a linear one for $c_1 \gg c_2$,
because $\mathbf{g} \sim \mathbf{c}_1$  and $ \mathbf{c}'_1 =\mathbf{c} -
p ( \mathbf{c}_1 \cdot \mathbf{n} )\mathbf{n}$, and the
$\mathbf{c}_2$-integration can be carried out in \Ref{BE}. The resulting
expression is,
\be \label{I-asympt}
I_\infty (c|f)= -\beta_\sigma c^\nu [1 -{\cal K}_\sigma
c^{-b(\sigma+1)/2}] f(c),
\ee
and we obtain for the coefficient,
\be \label{K}
{\cal K}_\sigma = \textstyle{\left({\Gamma
(\frac{d+\sigma}{2})}/{\Gamma (\frac{d-1}{2})}\right)} \left
(({2}/{\beta b (1-q^2)})^{(\sigma+1)/2} \right)
\ee
as derived in Appendix B of Ref.\cite{JSP-1}. Note also that this
coefficient vanishes in one dimension.

\section{Asymptotics for stable NESS }
\label{sec3}

Exact closed forms of the scaling solution of \Ref{int}  are not known in
general, but the high energy tail may be computed accurately with the
method developed in Ref. \cite{JSP-1}. Restricting the analysis to standard
arguments where the asymptotic form of $I(c|f)$ is $I_\infty(c|f) \sim
-\beta_\sigma c^\nu$ (see e.g. \cite{stretch-tails}), an interesting
feature emerges \cite{JSP-1,EPL}: in the region of stability ($b>0$),  the
asymptotic solution of \Ref{int} has a stretched exponential form, $ f
\sim \exp[-c^b]$, with $b=\nu +1-\theta$, while in cases of marginal
stability ($b \to 0^+$), $f \sim c^{-a}$ is of power law type with an {\it
a priori} unknown exponent $a$. As expected, $b$ decreases when $\nu$
decreases, since a tail particle with velocity $c \gg 1$ suffers
collisions at a rate $c^\nu$. The slower the rate, the slower the particle
redistributes its energy over the thermal range $c \lesssim 1$, which
results in an increasingly overpopulated high energy tail. When $\nu$ is
further decreased such that $b$ changes sign, the tail is no longer able
to equilibrate with the thermal ``bulk'', and the system cannot sustain a
steady state. A similar intuitive picture may be developed with respect to
$\theta$ in the NF cases \cite{EPL}.

The leading behavior of $f(c)$ is a generalization of the one obtained for
inelastic hard spheres ($\nu =1$) and Maxwell molecules ($\nu =0$). While
we had analyzed in details the case of WN driving in \cite{JSP-1}, we
focus here on inelastic gases driven by NF, which include the Gaussian
thermostat, or equivalently, the Freely Cooling gas (FC: $\theta=1$). The
method allows us to calculate the sub-leading correction for $c \gg 1$
to $I_\infty(c|f)$. This in turn yields important sub-leading
multiplicative correction factors  to $f(c)$ of exponential and power law
type, i.e.
\be \label{f-asympt}
\ln f(c) \sim -\beta c^b +\beta' c^{b'} + \chi \ln c +\mathcal{O}(1),
\ee
where $ b>b'>0$. This expression is in fact an asymptotic expansion of
$\ln f(c)$.  The limiting corrections as $b' \to 0^+$ are already
contained in the exponent $\chi$. Moreover, as soon as $b'$ becomes
negative, the correction term becomes $c^{-|b'|} \ll \mathcal{O}(1)$, and
should be neglected for consistency. In the spirit of asymptotic
expansions we only look  for the sub-dominant correction $c^{b'}$ with
$b'>0$ and set $\chi=0$. Only if $b'=0$ do we look for terms of type $ \chi
\ln c$. The goal of this section is to calculate the exponents
$\{b,b',\chi\} $ explicitly, and to express the coefficients $\{
\beta,\beta'\}$ in terms of the moments $ \langle \langle g^{\nu+2}
\rangle\rangle$ and $ \langle c^{\theta +1} \rangle $. These moments can
be independently measured in the DSMC (Direct Simulation Monte Carlo)
method (see Ref. \cite{JSP-1}). The sub-leading approximation supposedly
extends the agreement of theoretical predictions with measured DSMC data
to smaller $c$-values.

We start with the {\it NF integral equation}, obtained from \Ref{int} by
replacing the collision operator $I$ by the  full asymptotic form
$I_\infty$ in \Ref{I-asympt}. The last form is the appropriate one for
exponentially bound functions $f(c)$. This yields
\be \label{int-FC-2}
- \mathcal{B}_\sigma c^\nu \{1-   {\cal K}_\sigma (d) c^{-\half
(\sigma+1)b }\} f= c^{\theta} f^\prime +(d+\theta -1) c^{\theta -1} f,
\ee
where the $c$-independent factors have been combined into,
\be \Label{B}
\mathcal{B}_\sigma = \frac{4\beta_\sigma \av{c^{\theta+1}}}{ \lambda_2
\dav{g^{\nu+2}}} = \frac{2(d+\sigma)\av{c^{\theta+1}}}{(1 +\sigma)pq
\dav{g^{\nu+2}}},
\ee
and $\lambda_2 = 2pq \beta_{\sigma+2}$ and \Ref{cos-av} have been used.
Here the constant $\mathcal{B}_\sigma$ depends on all three model
parameters $(\nu,\sigma,\theta)$, and contains averages with the unknown
weight $f(c)$.

The parameters in $f(c)$ can be obtained from the full integral equation
\Ref{int-FC-2} by substituting the ansatz \Ref{f-asympt}, applying the
derivative, equating leading and sub-leading powers of $c$, and recalling
the relation $b>b'>0$. To leading order we have
$b\beta c^{b+\theta -1 }= c^\nu {\cal B}_\sigma$, yielding
\be \label{b-beta}
b=\nu+1-\theta; \qquad \beta b=\mathcal{B}_\sigma.
\ee
The exponent $b$ is the same as the one found in the stability analysis in
\Ref{rate-v0-NF}. These results are largely generalizations of special
cases, existing in the literature for $\theta =\{0,1\} ;\nu=\{0,1\};
\sigma =\{1,\nu\}$, derived in \cite{stretch-tails,MS}.

The remaining terms with sub-leading powers of $c$ have respectively the
exponents $ E_1=b'+\theta-1, E_2 = \theta-1, E_3 = \nu - \half (\sigma+1)b
$. First consider the case $ \sigma=1$, where $E_3=\nu-b=\theta-1=E_2$. As
$E_1>E_2$ the coefficient $\beta' =0$, equating the coefficients of the
remaining terms then yields the second line of the equation \Ref{B-sub}
below.  If
$\sigma >1$, then $E_1>E_2>E_3$, and the coefficient of each power has to
vanish, yielding the first line below. If $ \sigma<1$ we obtain the sub-leading
term by matching the exponents $E_1=E_3$, yielding the third line below.
Lower order terms with exponents $E_2$ have to be neglected for
consistency, hence $\chi =0$. So, the sub-leading results for {\it NF
driving} are,
\ba \Label{B-sub}
\sigma>1:& \beta' =0, & \chi =\!1\!-\!d\! -\!\theta
\nn \sigma=1: & \beta'=0, & \!\chi=1 \!-\! d \!-\!\theta \!+\!\beta b\mathcal{K}_1(d)
\!= \!-\!\theta \!+ \!\frac{(d-1)q^2}{1-q^2} \nn \sigma <1: & \chi=0, & b'
=\textstyle{\frac{1}{2}} b(1-\sigma), \: \beta'b' = \beta b
\mathcal{K}_\sigma(d).
\ea
In {\it one dimension} the results simplify substantially.
The collision kernel in
the Boltzmann equation \Ref{BE} lacks the angular integration $\int_\bn$,
$\beta_\sigma =1$ in \Ref{cos-av}, and $ \mathcal{K}_1(d) =0$  in \Ref{K}
for all $\sigma$, implying $\beta'=0$, and $b'$ is irrelevant. Then the
large-$c$ behavior of the distribution function is,
\be \label{B-sub-1d}
f(c) \sim c^{-\theta} \exp[ -\beta c^b ] \qquad (d=1).
\ee
Other simplifications occur for special values of the parameters $\theta$
and $\nu $. For $\nu=0$ (Maxwell molecules) the coefficient in \Ref{B}
simplifies as $\dav{ g^{\nu+2}}=\dav{g^2}=d$.

Simplification also occur for a case of special interest, the {\it free
cooling system } or equivalently  the {\it Gaussian thermostat} $(\theta=1
)$, where $\av{c^{\theta+1}} =\av{c^2} = \textstyle{\frac{1}{2}}d$ on
account of \Ref{norm-f}. Here the exponents and coefficients are to
leading order (see \Ref{B}),
\be \label{b-beta-FC}
b=\nu , \quad \beta b= \mathcal{B}_\sigma =
\frac{(d+\sigma)d}{(1+\sigma)pq \dav{g^{\nu+2}}} ,
\ee
and in sub-leading order,
\ba \Label{B-sub-FC}
\sigma>1:& \beta' =0, & \chi =-d
\nn \sigma=1: & \beta'=0, & \chi
= \frac{dq^2-1}{1-q^2} \\ \sigma <1: & \chi=0, & b'
=\textstyle{\frac{1}{2}} b(1-\sigma), \:
\beta'b' = \beta b \mathcal{K}_\sigma (d) \nonumber.
\ea
For free cooling
($\theta=1$) at $d=1$ we have ${\cal K}_\sigma (1) =0$, hence $\beta'=0$,
yielding,
\be \label{1d-FC}
f(c) \sim (1/c) \exp[-\beta c^b] .
\ee
Further simplification occurs in freely cooling Maxwell models
($\nu\!=\!0,\: \theta\!=\!1$), where $\mathcal{B}_\sigma
=(d+\sigma)/[(1+\sigma)pq\:] $. This is a marginally stable case ($b=\nu
+1-\theta =0$), and will be discussed in the next section.

\begin{figure}[htb]
\centerline{
\includegraphics[angle=0,width=7cm]{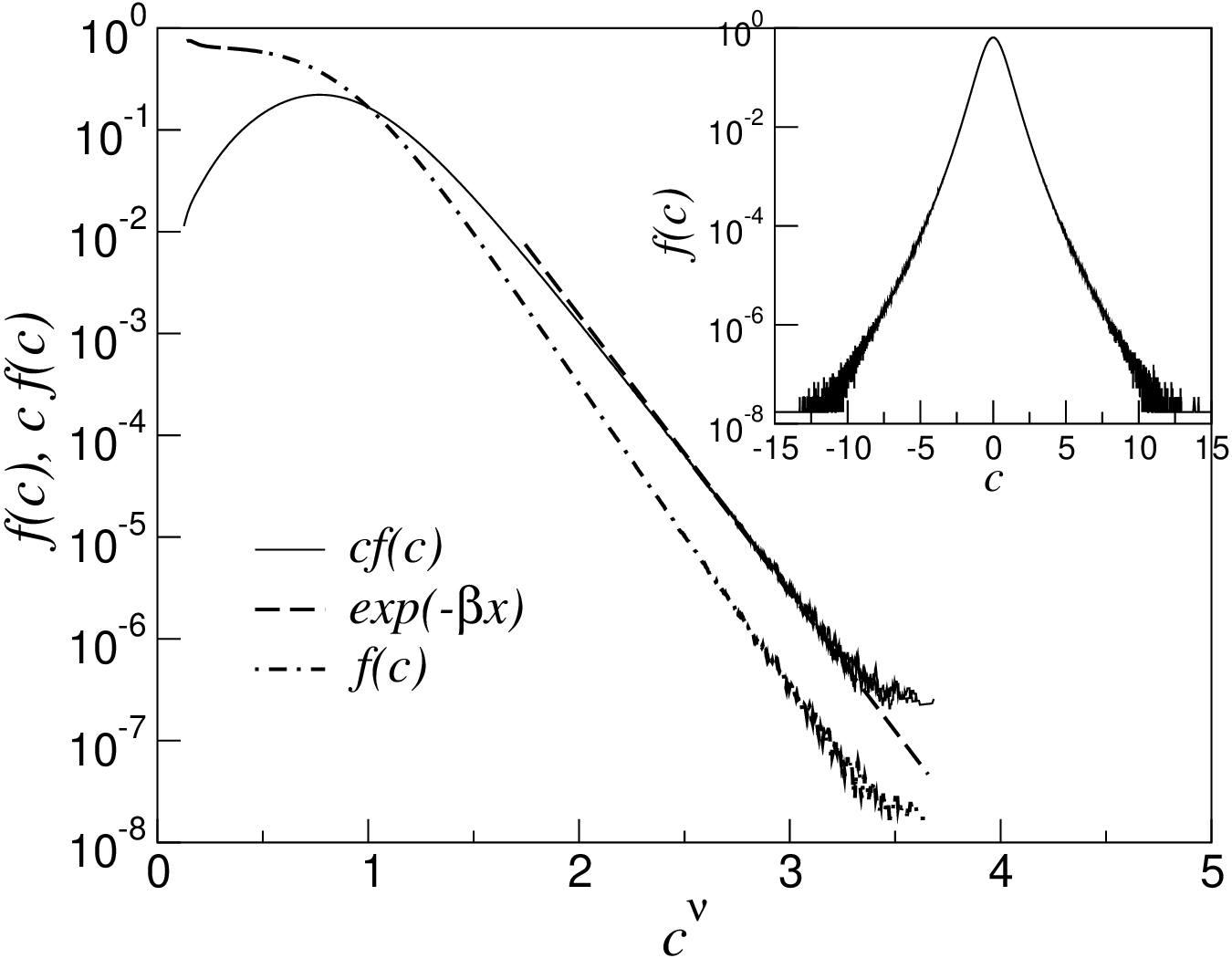} }
\caption{Free cooling with $(\theta =1, d=1, b=\nu=1/2, \alpha=0)
$, where $f(c)$ vs. $c^b$ and $c f(c)$ vs. $c^b$, are compared
with $\exp(-\beta c^b)$, to show the $\exp(-\beta c^b)/c$ behavior
of $f$. The solid line represents the Monte Carlo (DSMC) data. The
{\it inset} shows the overpopulation of the high energy tail when
compared to a Gaussian (on such a plot, a Gaussian would produce a
concave instead of convex graph).} \label{fig:FCd1nu.5}
\end{figure}

Finally we compare the analytic predictions with the  DSMC results. The
DSMC method offers a particularly efficient algorithm to solve the
nonlinear Boltzmann equation \cite{Bird}. Figure \ref{fig:FCd1nu.5}
shows for the one-dimensional case the simulation results (solid line) for
free cooling ($\theta=1 $) in the soft sphere model ($\nu = \half$) with
completely inelastic collisions ($p=q=\half$), compared with the analytic
results in zeroth approximation (dashed-dotted line), i.e. $f \sim \exp[
-\beta \sqrt{c}]$ in \Ref{b-beta-FC}, and in first approximation (dashed
line),  $ c f \sim \exp[ -\beta \sqrt{c}]$ in \Ref{1d-FC}, where $\beta =
8/\av{\av{g^{5/2}}}$ (according to \Ref{b-beta-FC}) is obtained by an
independent DSMC measurement of the two-particle moment. The zeroth
approximation has an effective slope {\it different} from the slope of the
first approximation. The latter essentially coincides with the DSMC
measurements for all $c \gtrsim 1.7$. Also note that the theoretical
curves can be shifted in the vertical direction to give the best possible
fit with the DSMC data, because the overall constant factor $\exp[{\cal
O}(1)]$ in $f(c)$ cannot be determined in our asymptotic analysis.

\begin{figure}[htb]
\vskip .5cm
\includegraphics[angle=0,width=7cm]{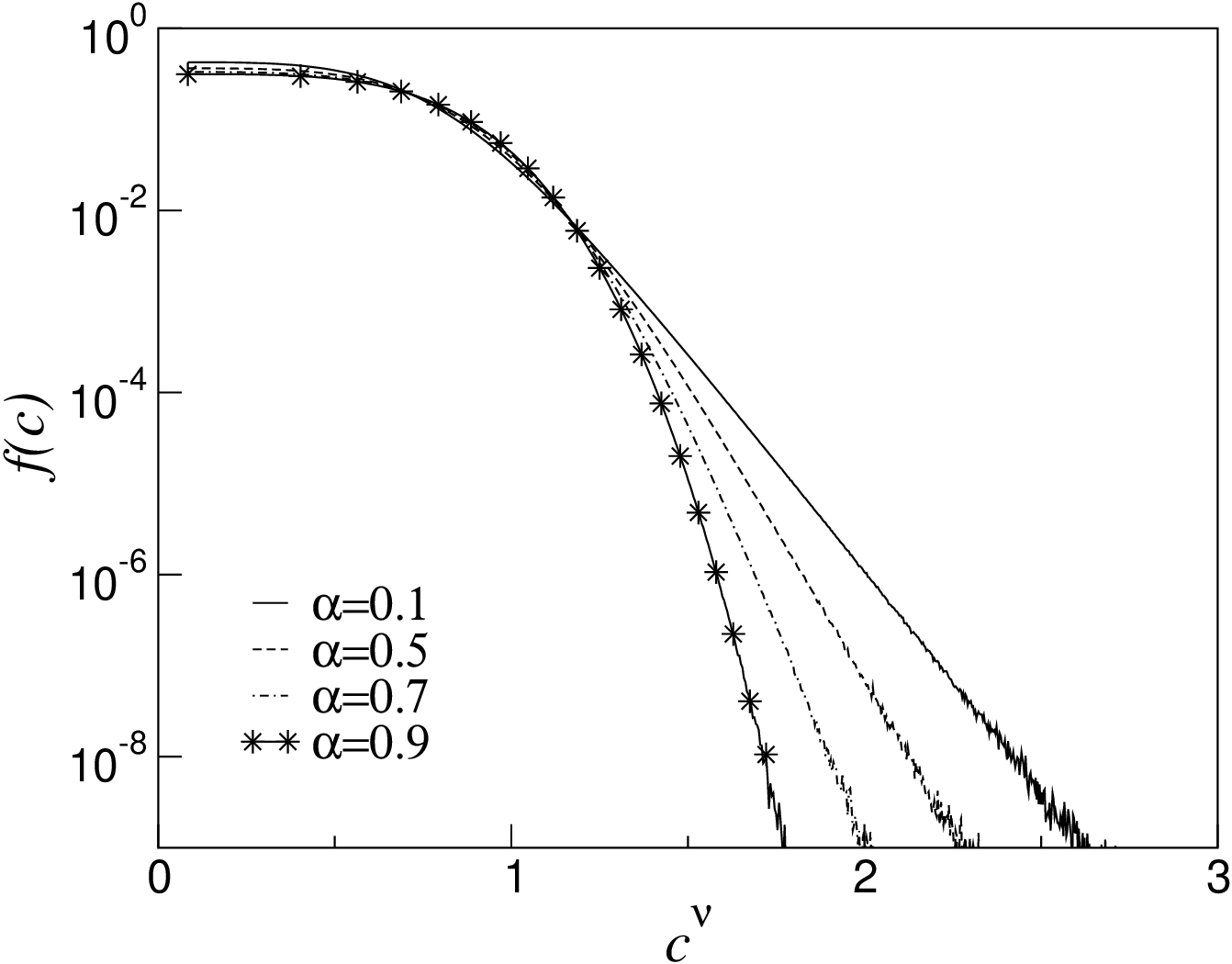}
\caption{Free cooling with $(\theta=1, d=2, b=\nu=\sigma= 0.5 )$ at
various values of $\alpha$. }
\label{fig:FCd2nu.5}
\end{figure}

The DSMC data in Fig. \ref{fig:FCd2nu.5} at large velocities show the
stretched Gaussian behavior $\exp[-\beta \sqrt{c}\,]$ for two dimensional
free cooling in the soft sphere model with ($b=\sigma=\nu=\half$). They
indicate that the coefficient $\beta$ increases with $\alpha$.
This figure
illustrates the overpopulation of the tail with respect to a Gaussian
(indistinguishable from the $\alpha=0.9$ curve here, shown with stars).

\begin{figure}[thb]
\includegraphics[width=7cm,angle=0]{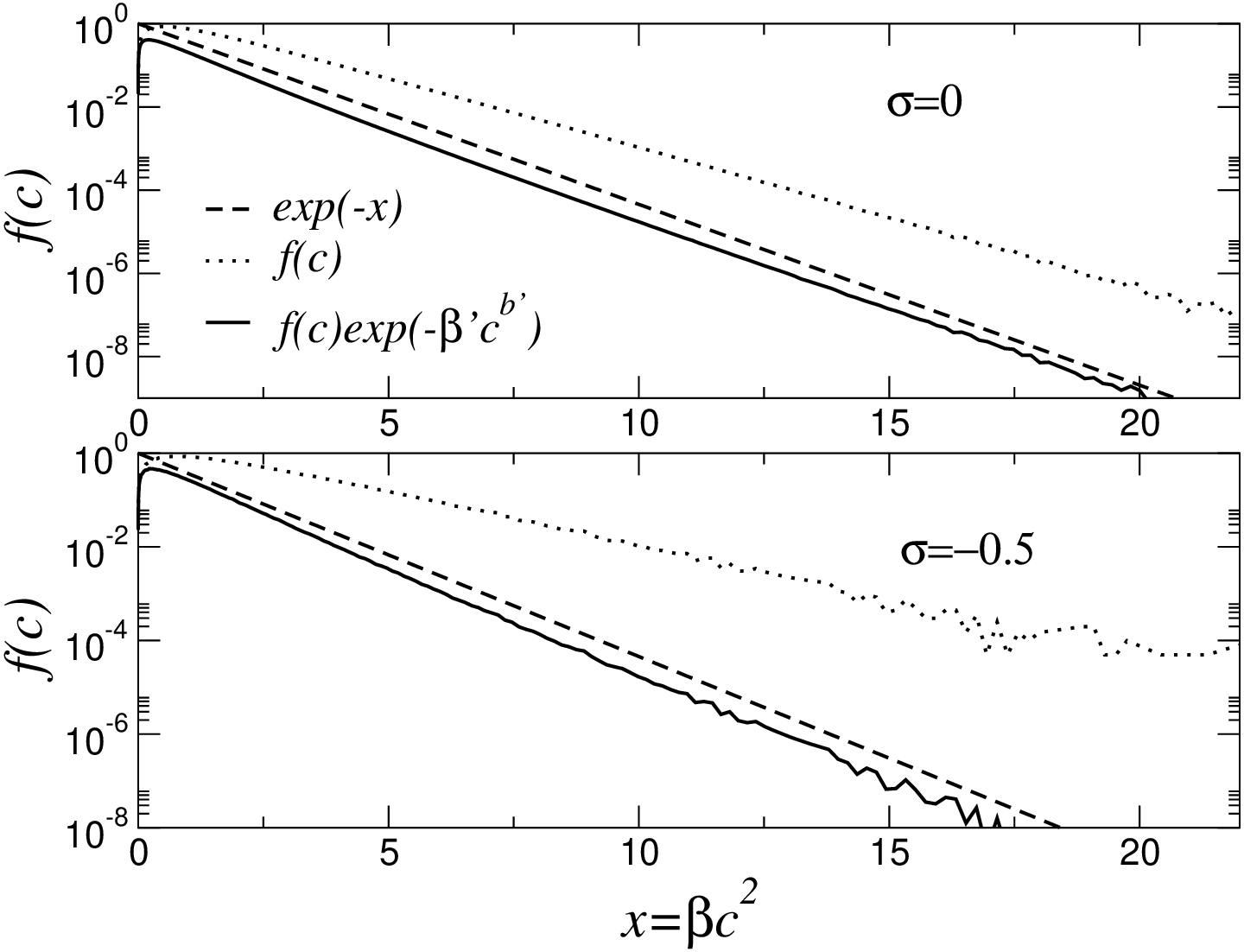}
\vskip -4.5mm \caption{Comparison of the velocity distribution
obtained from Monte Carlo (DSMC) simulations with the asymptotic
predictions. For the Gaussian thermostat ($\theta=1, d=2,
\alpha=0, \nu= 2)$ at $\sigma=0$ ({\it top}) and $\sigma=-0.5$
({\it bottom}) the exponents are $(b,b',\chi)= (2,1,0)$ and
$(2,1.5,0)$ respectively. DSMC data are plotted as $f(c)$ (dotted
line) and $\exp[-\beta' c^{b'}] f(c)$ (solid line) vs $x= \beta
c^b$, and compared with the theoretical prediction $e^{-x}$
(dashed line). Here $(\beta,\beta')\simeq (1.087,1.359)$ for
$\sigma=0$, and $\simeq (1.585,1.616)$ for $\sigma=-0.5$ have been
measured in the DSMC simulations from their definition given in
the text.} \label{fig:subleading}
\end{figure}

Striking examples of the importance of sub-leading corrections are shown
in Fig. \ref{fig:subleading}, for a two-dimensional model with $\theta=1$,
$\nu=2$, $\sigma=0$ and $\sigma=-0.5$. In these cases $b=2$, $b'=1$
(for $\sigma=0$) and $b'=3/2$ (for $\sigma=-0.5$). Comparison of the "raw"
DSMC data (dashed curve) with the dominant asymptotic prediction
$\exp(-\beta c^b)$ (dotted curve) shows no agreement. The reason is that
the simulated $c-$values are not large enough. However, the solid curve
(transformed DSMC data) $f(c)\exp[-\beta' c^{b'}]$ shows a striking
agreement with the theory $\exp(-\beta c^b)$, and demonstrates that the
sub-leading corrections extend the validity of the asymptotic theory to
much smaller $c-$values, thus enabling us to test the validity of theory,
and establish the importance of the sub-leading corrections. Striking
is the fact that such plots of $f(c)$ vs $c^b$ produces linear high energy
tails (in spite of the importance of the sub-leading correction), which
would then be well fitted with an effective value of $\beta$: $f(c)\sim
\exp(-\beta_{\text{eff}}\, c^b)$ (this is also the case in
Fig.~\ref{fig:FCd1nu.5}). As shown here, such an effective value can be
markedly different from the true $\beta$, which indicates that any fitting
procedure, aiming at computing $\beta$, is doomed to fail.

\begin{figure}[thb]
\vskip .65cm
\includegraphics[angle=0,width=7cm]{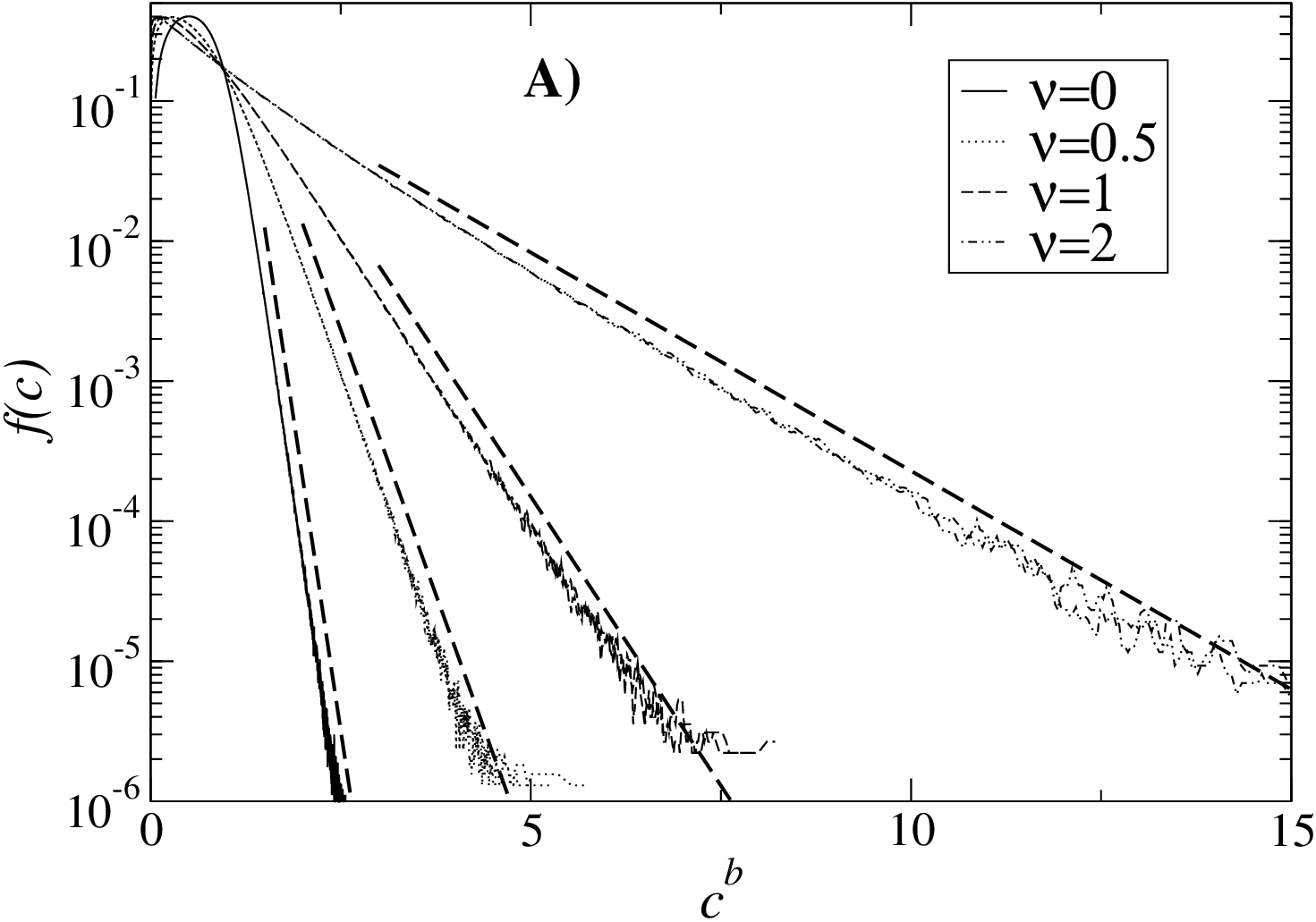}
\vskip .7 cm
\includegraphics[angle=0,width=7cm]{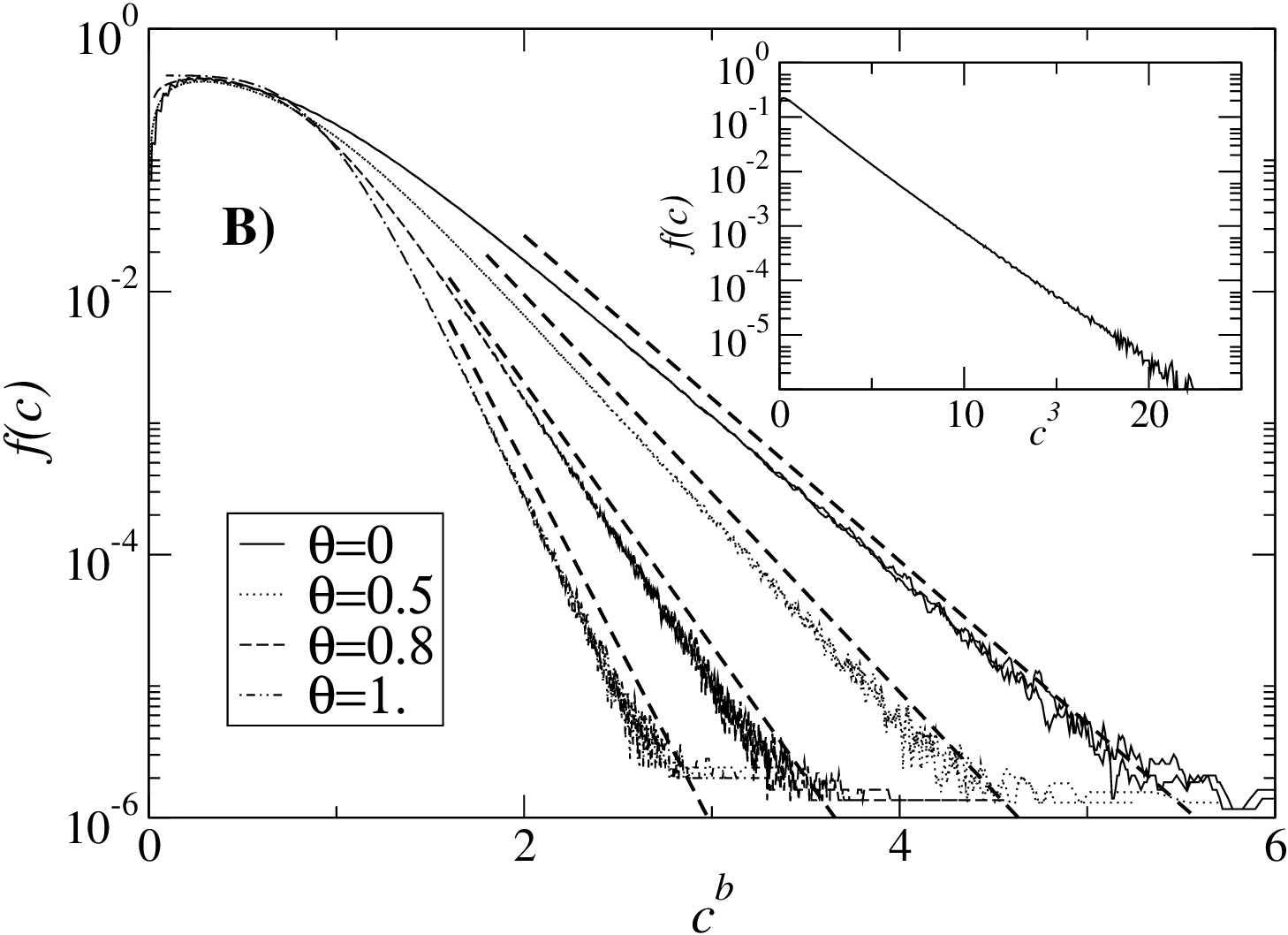}
\caption{ Negative friction with ($b=\nu+1-\theta, d=1, \alpha=0$). Plots
show $f(c)$ vs. $c^b$ (A) for various values of $\nu$ at
$\theta=0.5$, and (B) for various values of $\theta$ at
$\nu=0.5$. The dashed lines correspond to the predictions $\exp(-\beta
c^b)$ vs. $c^b$ with $\beta$ calculated in each case by DSMC. The {\it
inset} corresponds to a two-dimensional case ($\theta=0, \nu=2,
\alpha=0$), showing  $f(c)$ vs. $c^b$. } \label{figNF}
\end{figure}

Figure \ref{figNF} shows DSMC data for NF forcing with various values of
$\nu$ and $\theta$. The simulations confirm the large-$c$ predictions,
i.e. $\ln f(c) \sim - \beta c^b$, where $ b= \nu +1 -\theta$. Moreover,
the dashed lines show the agreement with the prediction $\exp(-\beta c^b)$
where the coefficient $\beta =2\av{|c|^{\theta+1}}/[bpq \dav{g^{\nu+2}}]$.
For those parameters, the sub-leading corrections are negligible.

Regarding the soft sphere systems in stable NESS ($b>0$), either freely
cooling or driven by Gaussian thermostats, we may conclude that the
agreement between analytic and DSMC results for high energy tails is very
good.

\section{Marginal stability, power law tails}
\label{sec4}

We now analyze the integral equation (\ref{int}) for the threshold models
($b=0$; this fixes the exponent $\nu$ at the threshold). Marginal
stability is a limiting property of a stable NESS as $b \to 0^+$, which
occurs in states,  {\it driven} either by white noise (see
Ref.\cite{JSP-1}) or by negative friction.

As we have seen in the previous section, the high energy tails for stable
states ($b>0$) have the generic form $f(c) \sim \exp[-\beta c^b]$ with
$\beta= \mathcal{B}_\sigma /b$, and sub-leading correction factors of
similar structure.  To illustrate how power law tails arise, we take the
limit of $f(c)$ as $b \to 0^+$ using the relation $(c^b-1)/b \sim \ln c$.
The result is,
\be \Label{S1}
f(c) \sim \lim_{b \to 0^+} c^\chi \exp [-\mathcal{B}_\sigma c^b/b] \equiv
c^{\chi-\eta}.
\ee
Of course $(\chi-{\eta})$ is not the full exponent of the tail, because
the exponential form above represents only the leading asymptotic behavior
for $b>0$. For instance, any correction factor $\exp[-\beta' c^{b'}]$,
where $b'=H(b) \to 0$ as $b \to 0$, gives additional contributions to
\Ref{S1}.

\subsection{Gaussian thermostat (NF: $\theta=1; b=\nu=0$)}

Here the {\it Maxwell molecules} are the marginally stable model. To
determine the full exponent of the power law tail we linearize the
nonlinear integral equation \Ref{int} at the stability threshold around
the "thermal bulk part" of $f(c)$, using \Ref{d+h} and \Ref{Lambda}. We
start with the simplest case of inelastic soft spheres, driven by a linear
friction force ($\theta =1$).

\begin{figure}
\vskip .65cm
\includegraphics[angle=0,width=7cm]{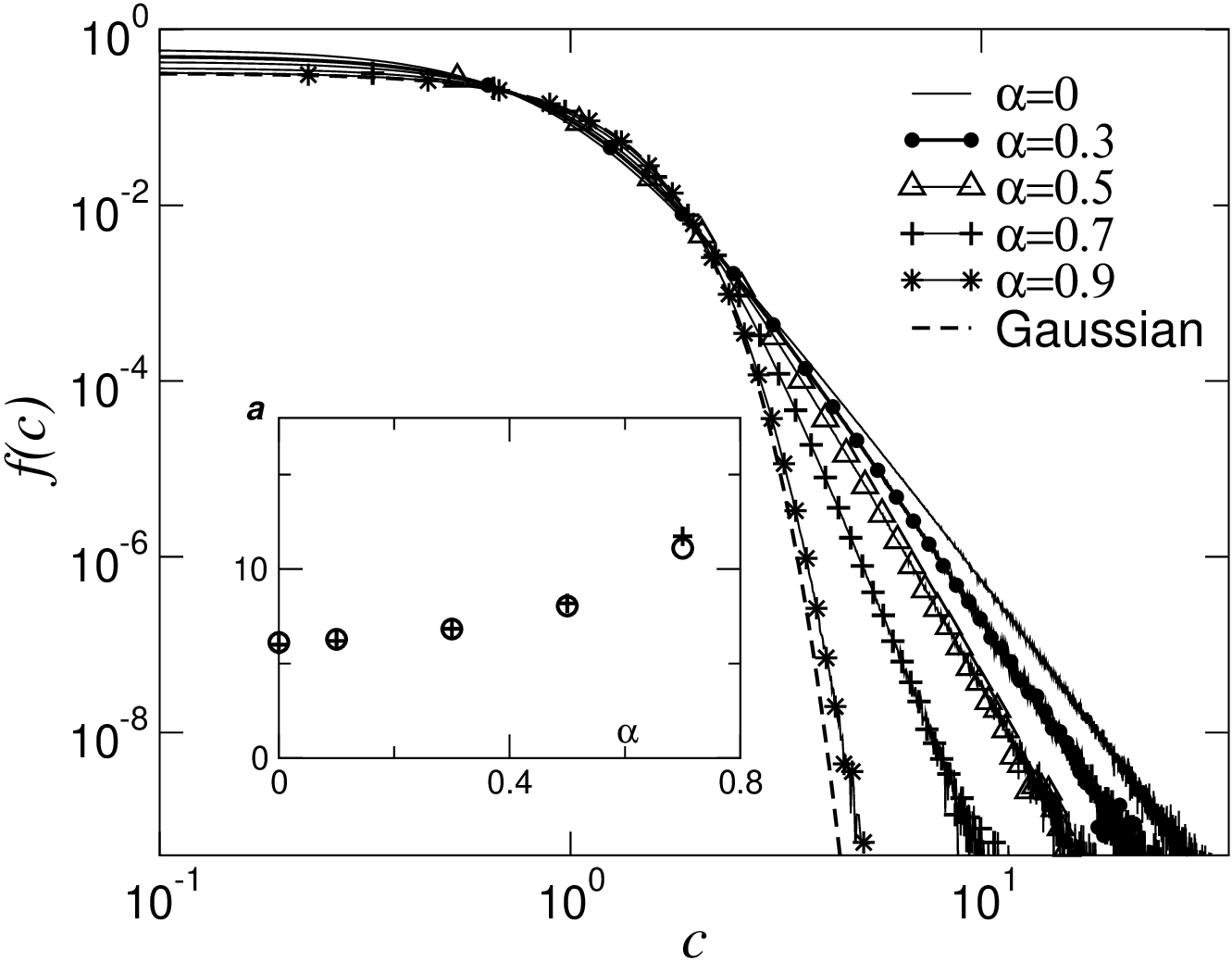}
\caption{Power law tails in  free cooling, obtained for the threshold
model $(\theta=1, d=2, \sigma=1, b=\nu=0 )$.
The {\it inset} compares predicted and measured exponents. As $\alpha$
increases the exponent increases, and the curve tends to a Gaussian. }
\label{fig_d2_free_nu0}
\end{figure}

Substitution of  $f(c) = \delta (\bc) +h(c)$ in the collision kernel of
\Ref{int} yields to linear order in $h(c)$, $I(c|\delta+h) = - \Lambda
h(c)$. The r.h.s. of \Ref{int} also simplifies, as $\dav{g^{\nu +2}} =
\dav{g^2} =d$, and the resulting integral equation is,
\be \Label{S4}
\Lambda h = - \half \lambda_2 \partial_\bc \cdot (\bc f) .
\ee
Inspection of this equation and \Ref{eigen-f} shows that the operators on
left and right hand side, when acting on the right eigenfunction
$1/c^{s+d}$  (recall that $\nu=0$) generate new powers of $c$. Solving the
integral equation implies that one determines the value $s^*$ that makes
both exponents equal, leading to the transcendental equation,
\be \Label{TE-FC}
\lambda_s = \half s \lambda_2 = spq \beta_{\sigma+2}.
\ee
Consequently, the solution of \Ref{S4}, which presents the asymptotic
large-$c$ solution of \Ref{int}, is the power law tail,
\be \Label{S6}
f(c) \sim h(c) \sim 1/c^{s^*+d} \qquad (c \gg 1) .
\ee
If the transcendental equation has more solutions, then the largest root
$s^*$ is the relevant one, because the energy $\av{c^2}$, and possible
moments $\av{c^a}$ and $\dav{g^a}$, appearing in the transcendental
equations (see next subsection) must be finite, imposing $s^*>\max \{2,a
\}$. So, the obvious solution of \Ref{TE-FC},  $s^*=2$, has to be
rejected. However the equation has a second solution with $s^*>2$, because
$\lambda_s$ is a concave function of $s$. This can be seen directly from a
graphical solution by adding in Fig.~\ref{figlambdas} the line $y(s) =
\half  s \lambda_2$. The numerical values of $s^*(\alpha)$, obtained from
the numerical solution of \Ref{TE-FC}, are shown in the inset of
Fig. \ref{fig_d2_free_nu0}
for the two-dimensional system. The main plot shows the comparison of the
DSMC measurements of $f(c)$ for this system compared to the theoretical
predictions.

It is also instructive to consider the one-dimensional version of
\Ref{TE-FC}, which can be solved analytically. Then the eigenvalue
\Ref{eigen-val} simplifies to $\lambda_s =1-q^s-p^s$, and \Ref{TE-FC}
becomes, $1-q^s-p^s = spq$,  with solutions $s^* =\{2,3\}$, and $s^*=3$ is
the relevant one, and  $f(c) \sim 1/c^{s^*+d} \sim 1/c^4$ in agreement
with the exact solution $ f(c) =(2/\pi)/[1+c^2]^2$,  found in \cite{Rome}.

For $d>1$ the transcendental equation  can not be solved analytically,
except in a few limiting cases, that we consider first. In the {\it
elastic limit} ($\alpha \to 1$ or $q \to 0$) one only finds a meaningful
solution of \Ref{TE-FC} by letting simultaneously $s \to \infty$ while
keeping $sq = \xi=$ fixed. As $\beta^{\sigma+2} /\beta^\sigma =(\sigma
+1)/(\sigma +d)$, and $\lambda_s/\beta^\sigma\to 1$ as $s \to \infty$ (see
Fig. \ref{figlambdas} or Eq.(3.12) in Ref.\cite{JSP-1}), the transcendental
equation \Ref{TE-NF} approaches  $1 \simeq \xi
(1+\sigma)/(d+\sigma)$, yielding the solution,
\be \Label{S7}
s^*_\sigma=\xi/q  \sim \left[({d+\sigma})/({1+\sigma})\right]/q \qquad
(\alpha \to 1)
\ee
for general $\sigma$. In the elastic limit as $\alpha=1-2q \to 1$, the
root $s^*_\sigma$ moves to \,$\infty$\,  and the algebraic tail
disappears, as required by consistency with the Maxwell distribution in
the elastic limit. Using the large $s-$expansion of \Ref{TE-FC} it is
straightforward to obtain additional sub-leading corrections.

Another case where the integral equation \Ref{TE-FC} can be solved
analytically is at {\it large dimensions} \cite{KBN02}. To do so it is
convenient to divide \Ref{TE-FC} by $\beta_\sigma$. As $ d \to \infty$ its
right hand side approaches $ spq (1+\sigma)/d$. So, one finds only a
meaningful solution by simultaneously letting $s \to \infty$ while keeping
$x=s/d=$ fixed. To calculate $\lambda_s/\beta_\sigma$ from \Ref{eigen-val}
in this {\it coupled} limit we use the relation,
\be \label{S7a}
\lim_{d \to \infty}  {_2}F_1(\textstyle{-\frac{xd}{2},
\frac{\sigma+1}{2};\frac{\sigma+d}{2}}|z) = \sum_{n=0}^\infty (\half)_n
(-xz)^n = \frac{1}{\sqrt{1+zx}},
\ee
where $(a)_n \equiv \Gamma (a+n)/\Gamma (a)$. This relation can be derived
starting from the Gauss hyper-geometric series \cite{Abram+Stegun} for
$_{2}F_1 (a,b;c|z)$ by taking the $(d \to \infty)$ limit term by term, and
subsequently using the relation $_{2}F_1 (a,b;b|z) = (1-z)^{-a}$. Then
\Ref{TE-FC} for the present threshold model simplifies to,
\be \Label{S8}
1-\left( 1+x(1-q^2)\right)^{-(1+ \sigma)/2} =x pq(1+\sigma).
\ee
For the $\sigma$-values, mostly considered in the literature, i.e. the
model with $(\sigma =1)$ \cite{Math,EB-JSP}, and the
mathematically convenient model $(\sigma=0)$ \cite{Maxwell,EB-JSP}, the
above equation can be solved analytically. For the Maxwell model  with
$\sigma=1$ it is a quadratic equation, and for $\sigma=0$ it is a cubic
equation. The root $x=0$ is not a solution  of \Ref{TE-FC} because
$\lambda_s$ in \Ref{eigen-val} holds only for $s>0$. The resulting
$s^*_\sigma$ in \Ref{S6} becomes in the coupled limit $d\to \infty, s \to
\infty$ with $s/d =x =$ fixed,
\be \Label{S9}
s^*_\sigma = x^*_\sigma d \simeq \left\{ \begin{array}{ll}
\frac{d}{2q(1+q)} & \quad (\sigma=1)\\ [2mm] \frac{d}{q(1-q^2)} [1+\half q
+\sqrt{q(1+ \textstyle{\frac{5}{4}q})}]& \quad (\sigma=0)
\end{array} \right..
\ee
The exponent $s^*_\sigma+d$, obtained here, disagrees
with the result of Ref. \cite{KBN02} in the sign in front of the square
root.
We note that the $\alpha-$dependence of $s^*_0$ and $s^*_1$ in the last
equation  is somewhat different  at large $d$. The exponents in \Ref{S7}
and \Ref{S9} agree in the respective limits $d \to \infty$ and $\alpha \to
1$.

\begin{figure}
\includegraphics[angle=0,width=7cm]{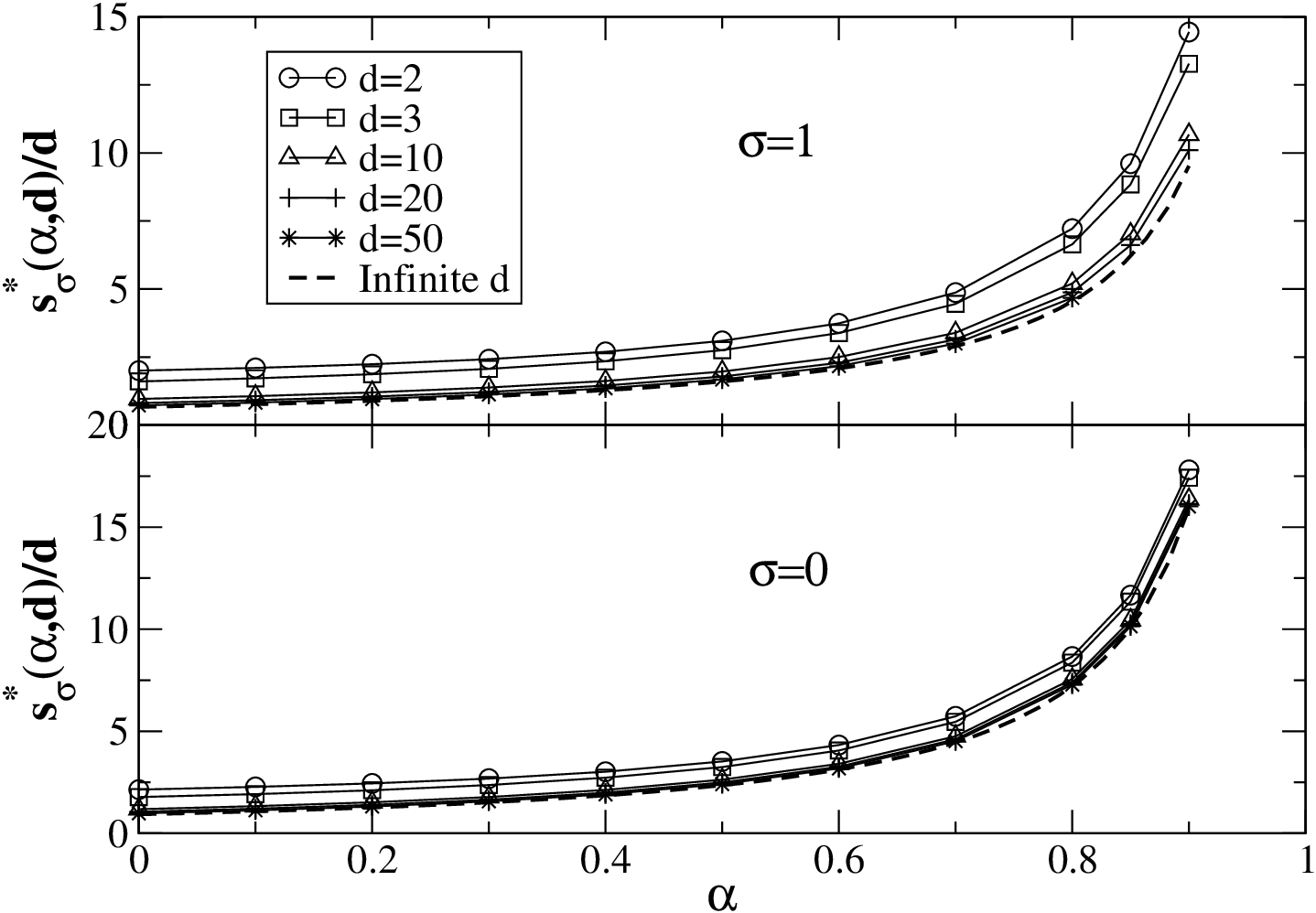}
\caption{ $s^*_\sigma(\alpha,d)/d$  vs $\alpha$ for the two variants of
$d-$dimensional Maxwell models, i.e. ({\it top}) with $(\sigma=1,\nu=0)$
and ({\it bottom}) with $(\sigma=\nu=0)$. The dashed lines correspond to
the analytic results \Ref{S9} for large $d$.} \label{fig_expvsalpha}
\end{figure}

Equation \Ref{TE-FC} can easily be solved {\it numerically}.
For the Maxwell model with $\sigma=1$ the resulting exponents $s^*_1$ and
$s^*_0$ as a function of $\alpha$ for various $d$ are plotted as $s^*_1/d$
and $s^*_0/d$ in Fig. \ref{fig_expvsalpha}. As shown in the inset of
Fig. \ref{fig_d2_free_nu0}, the agreement with DSMC simulations is very
good.

Most results of this subsection, applying to Maxwell models ($\nu=0$),
have been derived already in the literature using an entirely different
mention, namely by Fourier transformation with respect to the velocity
variables.  The Fourier transform method can only be applied to Maxwell
models ($\nu=0)$ where the microscopic collision frequency is independent
of the relative velocity $g$, leading to a collision kernel $I(c|f)$ that
is a convolution product in velocity space. It simplifies to an ordinary
product after Fourier transformation. The method can not be generalized to
inelastic soft sphere models with $\nu \neq 0$. Our method on the
other hand can be applied for all values of $\nu$.

\subsection{Nonlinear negative friction\\
(NF: $\theta \geq 0; b=\nu+1-\theta=0$)}

In this case, the threshold model is the soft sphere mode with $b=0$ or $\nu
=\theta -1$.
The corresponding scaling equation for the high
energy tail is obtained by setting $ \nu=\theta -1$ in \Ref{int}, and
reads,
 \be \Label{S19}
 I(c|\delta +h) =-\Lambda h = \half
\lambda_2\Gamma(\theta) {\bf \partial} \cdot ( \hat{\bc} c^\theta h),
\ee
where we have defined the ratio of the moments $\Gamma (\theta)$ as,
\be \Label{S20}
\Gamma(\theta) \equiv \frac{\dav{g^{\theta+1}}}{2\av{c^{\theta+1}}}.
\ee
This quantity should not be confused with the Euler Gamma function.
We also note that $\Gamma(\theta)$ is {\it unknown} a priori, as it
depends on the full unknown scaling form $f(c)$ with $c \in (0,\infty)$.
Inspection of \Ref{S19} shows again that the operators on left and right
hand side of \Ref{S19}, when acting on the right eigenfunction
$1/c^{s+d+\nu}$ with $\nu = \theta-1$, will produce new powers of $c$, and
one determines the value $s^*$, that makes both exponents equal, by
solving the transcendental equation,
\be \Label{TE-NF}
\lambda_s = \half s \lambda_2 \Gamma(\theta)= spq
\beta_{\sigma+2}\Gamma(\theta).
\ee
We recall that $\lambda_s$ is the same
for all inelastic soft sphere models. We further note that
$\Gamma(\theta=1)=1$, as can be verified from \Ref{S20} and the
normalization $\av{c^2} =d/2$, and we recover the transcendental equation
\Ref{TE-FC} for linear friction.

Denoting the relevant root of \Ref{TE-NF} by $s^*_\sigma$ the solution of
\Ref{S19} is the right eigenfunction of $\Lambda$ with eigenvalue
$\lambda_{s^*_\sigma}$, i.e.
\be \Label{S21}
f(c) \sim h(c) \sim c^{-s^*_\sigma-d-\theta +1}.
\ee
So at the stability threshold  for driving by nonlinear friction ($\nu
=\theta-1$), there exists again  a power law tail in the scaling solution
of the Boltzmann equation for soft sphere models, provided \Ref{TE-NF}
does indeed have a real positive  solution.

\begin{figure}
\includegraphics[angle=0,width=7cm]{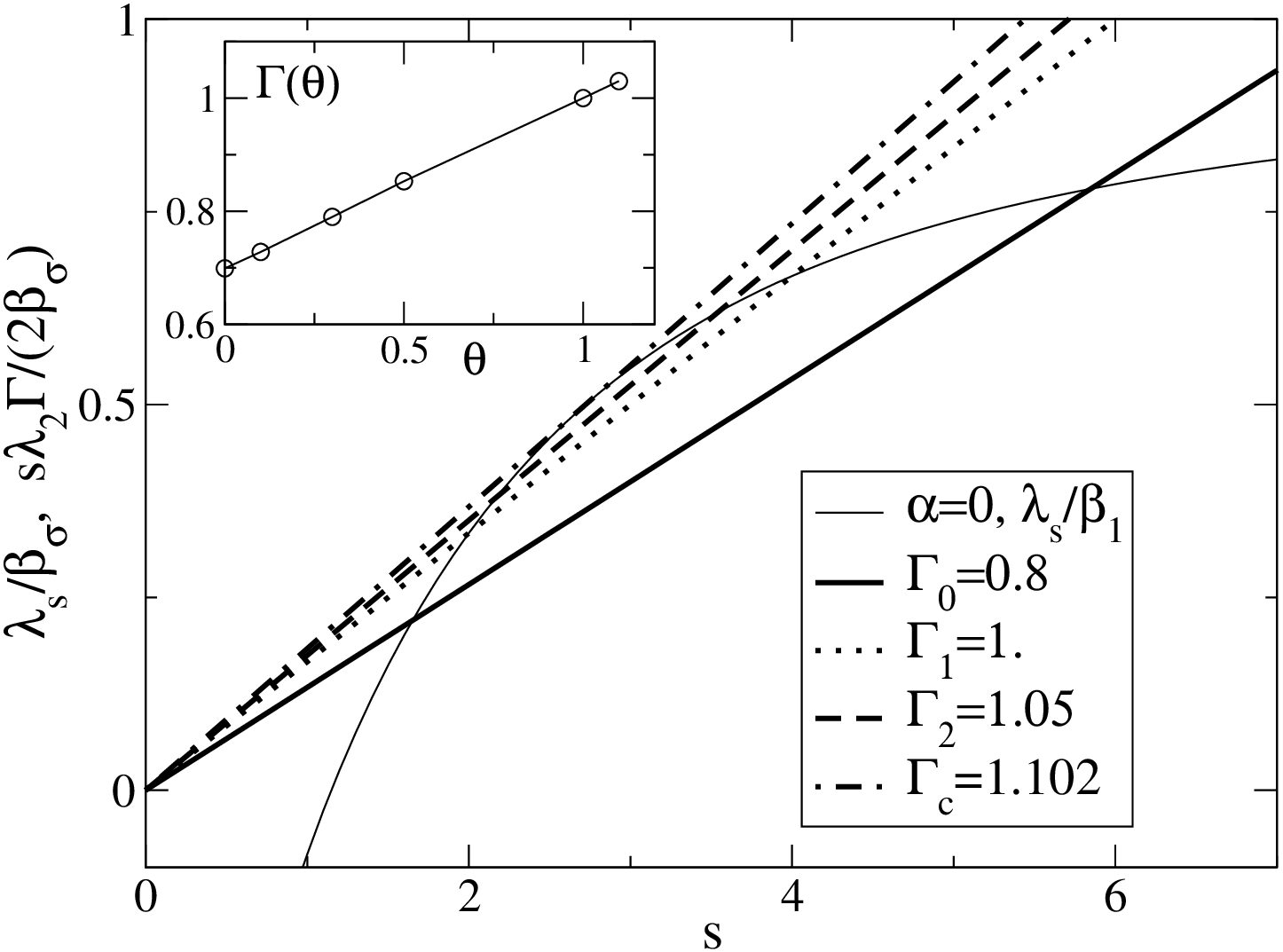} \caption
{NF at ($d=2$, $\sigma=1$, $\nu=\theta-1$, $\alpha=0 $). Graphical
solution of \protect{\Ref{TE-NF}} for the marginally stable NF
driven soft sphere model. The concave curve represents the
eigenvalue $y_1(s) = \lambda_s(1) /\beta_1 $ (solid line) and the
straight lines represent $y_2(s)$ (see text) for different values
of $\Gamma(\theta)$, labelled from bottom to top by $\Gamma_n
(n=0,1,2)$. The inset shows $\Gamma(\theta)$ versus $\theta$ as
obtained from DSMC measurements. The slope of the tangent line is
labeled by $\Gamma_c$. The largest intersection point corresponds
for a given value of $\Gamma(\theta)$ to the root $a_n(\sigma)$,
which determines the power law tail $f(c) \sim
1/c^{a+d+\theta-1}$.} \label{fig_solTE}
\end{figure}

Extracting the largest root from \Ref{TE-NF} is somewhat more complicated than
in equation \Ref{TE-FC}, because of the {\it unknown} factor $\Gamma(\theta)$.
Even for $d=1$ there are no simple exact solutions. To obtain $\Gamma(\theta)$
we determine the moments in \Ref{S20} and their ratio $\Gamma (\theta)$ by
direct DSMC measurements. The inset of Fig. \ref{fig_solTE} shows
$\Gamma(\theta)$ resulting from these measurements at $\alpha=0$ in two
dimensions.  The plot shows that $\Gamma(\theta)$ is an approximately linear
function increasing with $\theta$. At this point, it is noteworthy that a
Gaussian ansatz for the velocity distribution yields
$\Gamma(\theta)=2^{(\theta-1)/2}$. This provides an excellent approximation
(not shown) \footnote{We thank an anonymous referee for this remark.}, 
which also coincides with the exact value at $\theta =-1$.  There
the linear approximation is slightly off. However, in the physically relevant
range, $\theta \in [0,1]$, the linear approximation is slightly better.
The following analytical results confirm this
trend: $\Gamma =\half, 1$ for $\theta =-1,1$ respectively. The resulting
$\Gamma(\theta)$ is used as a known input parameter in \Ref{TE-NF}.

Once $\Gamma(\theta)$ is known from DSMC measurements, one can construct a
simple graphical method for solving \Ref{TE-NF} and classifying its
possible solutions for different values of $\theta$ and $\alpha$. Here we
discuss only the $\nu$ models with $\sigma=1$.
This is done by plotting in Fig. \ref{fig_solTE} for a fixed value of
$\alpha$ the curve, $y_1(s)=\lambda_s/\beta_1$ (l.h.s. of \Ref{TE-NF}),
and the straight lines, $y_2(s) =[2pq \Gamma(\theta)/(d+1)]s$ (r.h.s. of
\Ref{TE-NF}, as follows from
$\beta_3/\beta_1=2/(d+1)$), for
different values of $\theta$, and determine the largest
intersection point. Here $\Gamma_c $ defines the slope of the line, $y =
[2pq \Gamma_c/(d+1)] s $, through the origin, that is {\it tangent} to
curve $\lambda_s$. The largest intersection point of the eigenvalue curve
with the line, labelled $\Gamma(\theta=1)=1$, represents the graphical
solution for the linear friction case, and the relevant root $s^*(\alpha)$
has already been obtained in Fig. \ref{fig_d2_free_nu0}
for two dimensions \cite{EB-JSP,EB-SpringII}.
For the nonlinear case we obtain the following scenario. As $\theta $
decreases from 1 to 0, the ratio $\Gamma (\theta)$ decreases from 1 to
some value $\Gamma (0) > 1/2$, and the largest root $s^*_+ =
s^*_+(\theta,\alpha)$ grows  from $s^*$ to some value $s^*_+(0,\alpha)$.
As $\theta$ grows larger than 1, $\Gamma(\theta,\alpha)$ increases from 1
to $\Gamma_c(\alpha)$, and the largest root $s^*_- = s^*_-(\theta,\alpha)$
decrease from $s^*$ to $s^*_c$, as shown in Fig. \ref{fig_solTE}. For
$\Gamma (\theta)> \Gamma_c$ the root of the transcendental equation
becomes complex. The corresponding tail with an oscillatory pre-factor  is
no longer everywhere non-negative, and thus becomes unphysical.

\begin{figure}
\includegraphics[angle=0,width=7cm]{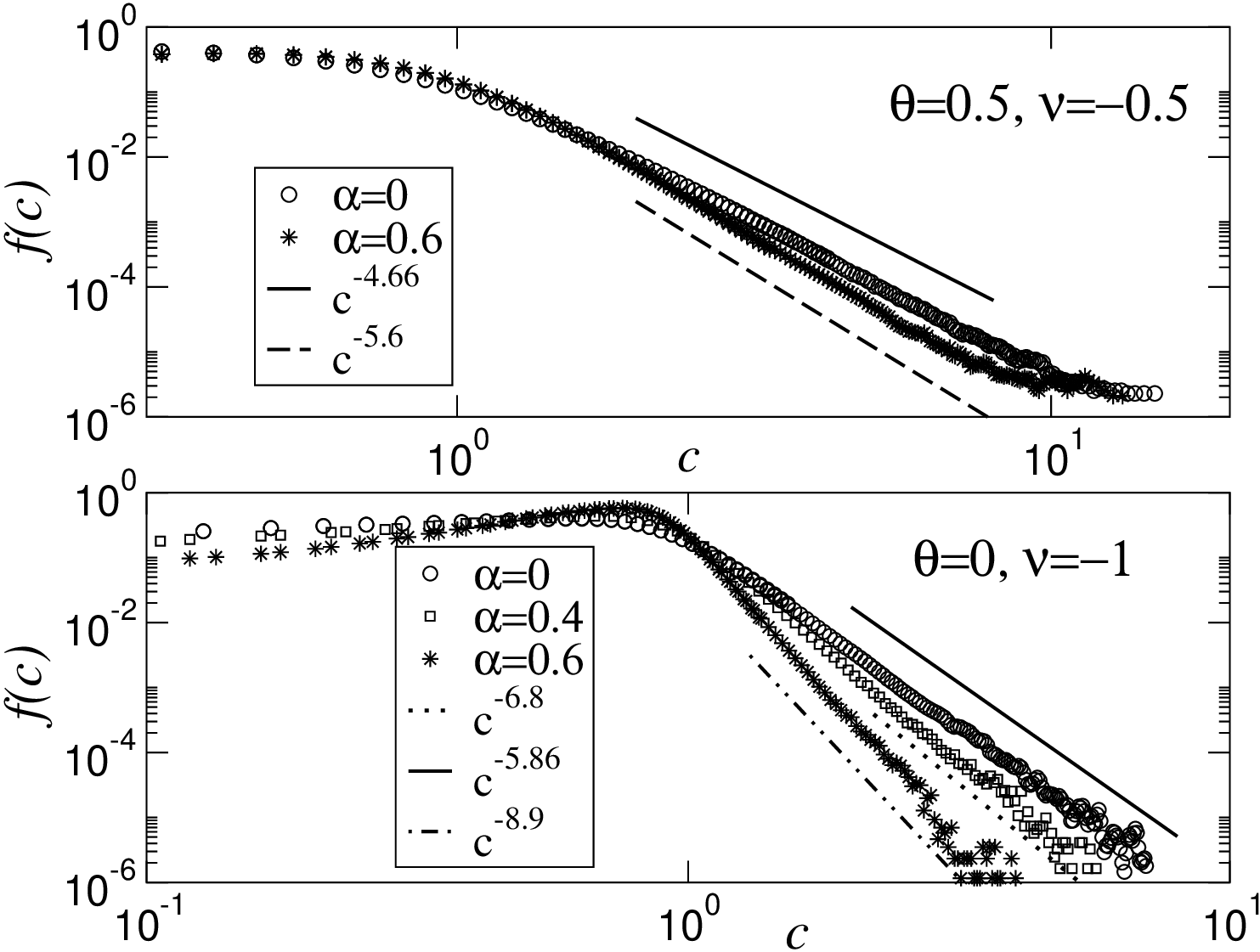}
\vskip .65cm
\caption { NF at stability threshold ($d=1, b=\nu+1-\theta=0$). ({\it
top}): $(\theta,\nu)=(0.5,-0.5)$; ({\it bottom}): $(\theta,\nu)=(0,-1)$.
The lines are the predicted power law tails, $f(c) \sim 1/c^{a+d+\theta
-1}$, following from the construction discussed in Fig.
\ref{fig_solTE}. } \label{fig4.11}
\end{figure}

\begin{figure}
\includegraphics[angle=0,width=7cm]{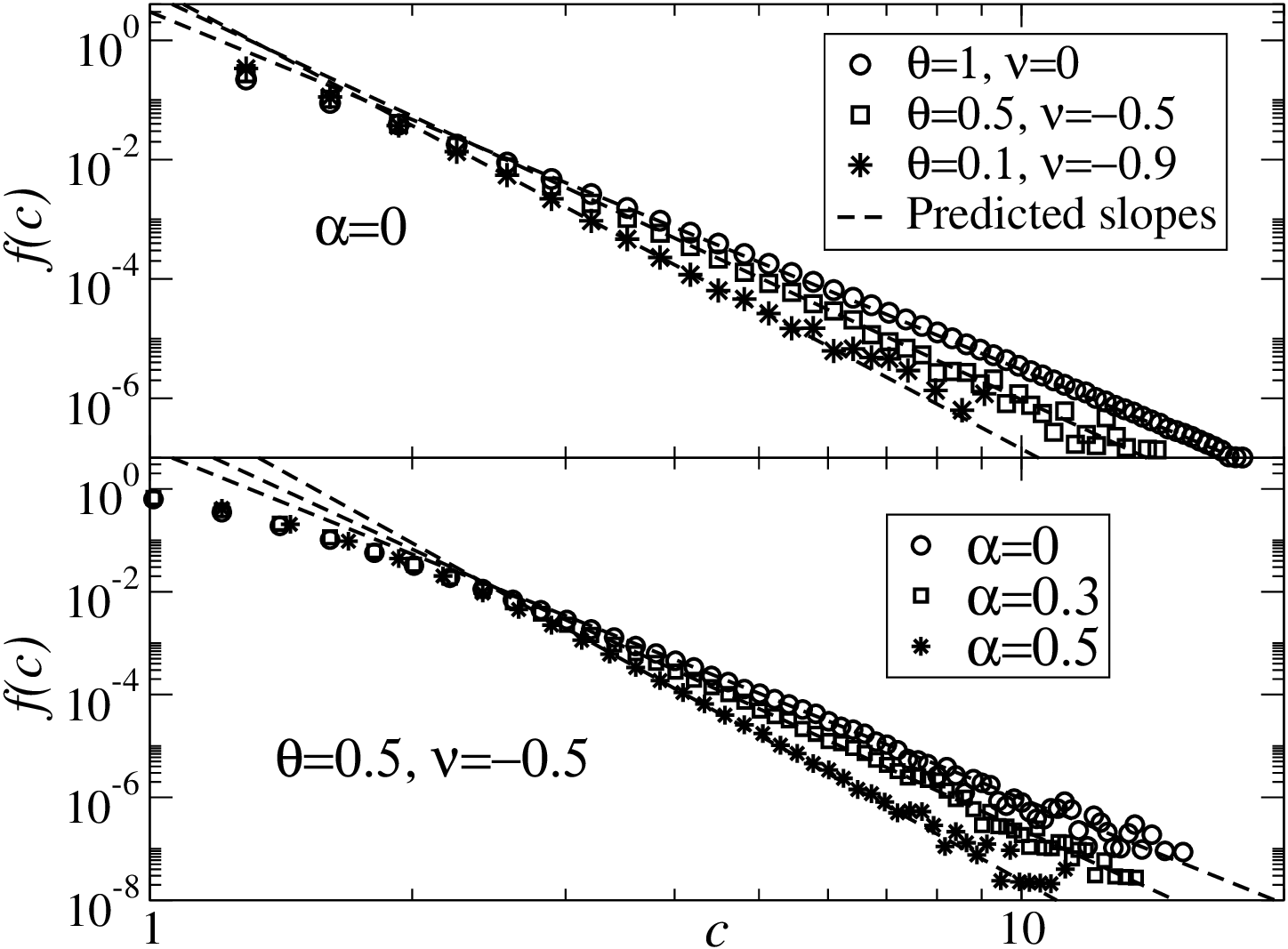}
\vskip .65cm
\vskip .65cm
\includegraphics[angle=0,width=7cm]{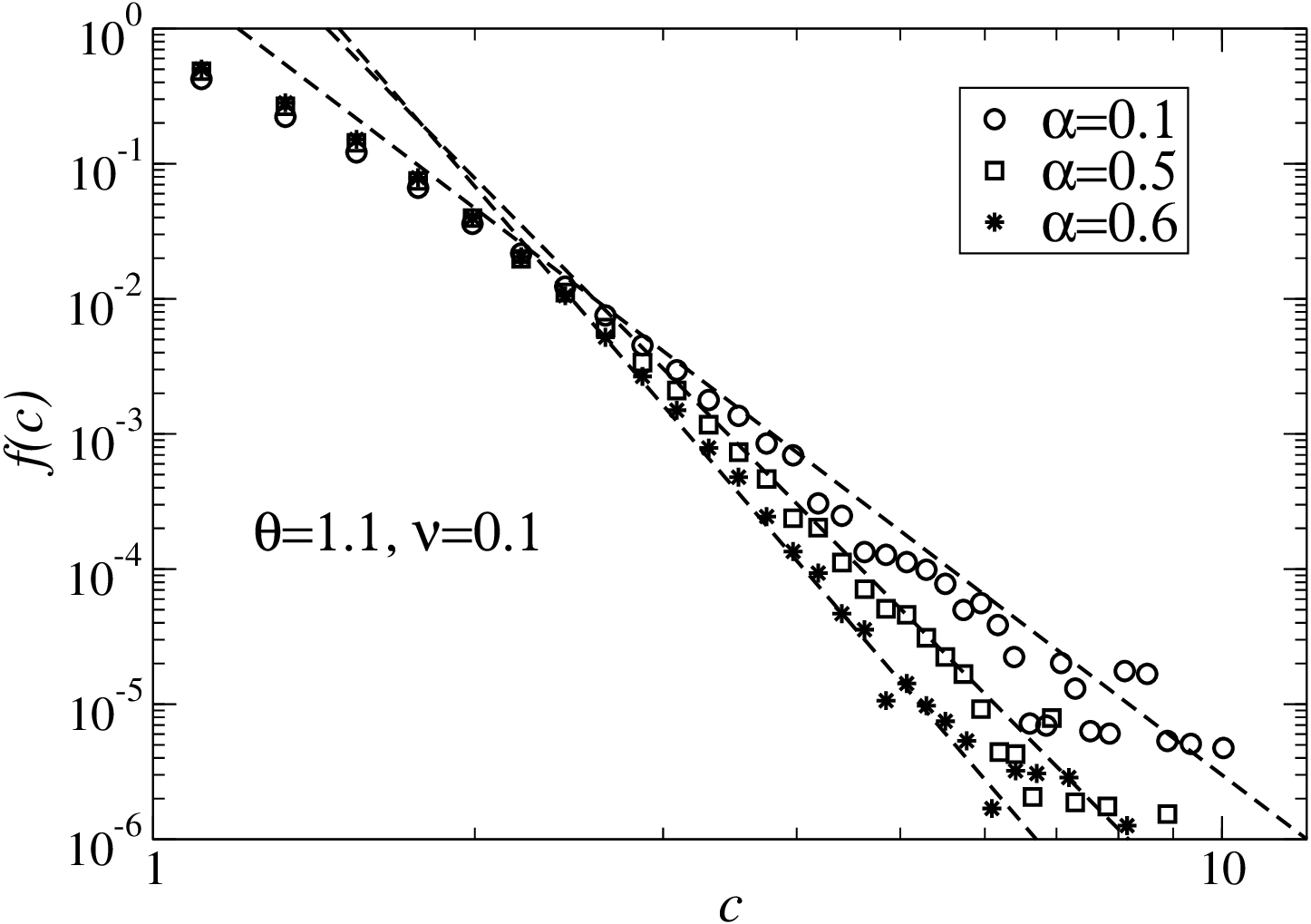}
\caption {NF at stability threshold ($d=2, b=\nu+1-\theta=0 $). ({\it
top}): $\alpha=0$ and $(\theta,\nu)=\{(1,0),(0.5,-0.5),(0.1,-0.9)\}$;
({\it middle}): $(\theta,\nu)=(0.5,-0.5)$ and $\alpha= \{0.0, 0.3, 0.5\}$;
({\it bottom}): $(\theta, \nu) = (1.1,0.1)$ and $\alpha =
\{0.1,0.5,0.6\}$. The plots show the predicted power law tails, $f(c) \sim
1/c^E$, as dashed lines with exponents $E= a+d+\theta -1$, where $a$ is
calculated from the transcendental equation \Ref{TE-NF} using the measured
values of $\Gamma(\theta)$ (see also Fig. \ref{fig_solTE}) . The predicted exponents are: ({\it top}): $ E
= \{6.0, 6.8, 7.75\}$ ; ({\it middle}): $ E = \{ 6.8, 7.8, 9.4\}$; ({\it
bottom}): $ E = \{6.0, 8.0, 9.2\}$. These exponents  show very good
agreement with DSMC data. }\label{fig4.10}
\end{figure}

We finally present a comparison of our analytical predictions with the
result of DSMC simulations for several parameter values in Figs.
\ref{fig4.11} and \ref{fig4.10}. Due to the marginally stable character of
the NESS, simulations are quite difficult and time-consuming.
Nevertheless, an excellent agreement is obtained for all parameter values.
Note that the values of the power-law exponents are large so that a direct
fit to power-law forms would not be very precise, and could not exclude
other fitting forms (since at most one decade in $c$ is covered).

\section{Inelastic BGK models}
\label{sec6}

In this section we study a simple inelastic BGK (Bhatnagar - Gross -
Krook) model for homogeneous velocity relaxation \cite{BMD,EB-SpringII},
which only takes the most essential features of the complex nonlinear
collision operator into account. The goal is to understand how much of the
rich behavior of the Boltzmann equation, described in the present paper
and in \cite{JSP-1}, is preserved in such a {\em linear} model. The
analytic results of the previous sections, and of Refs. \cite{JSP-1,EPL,QE},
are restricted to asymptotic solutions, which can be applied directly to
\Ref{a1}.
The BGK kinetic equations allow to go further, since they reduce to
simple linear first and second order inhomogeneous ODE's, which can be
solved {\it exactly}, at least for systems that are freely cooling, or
equivalently driven by linear negative friction,  as well as for systems
driven by white noise. Although the present paper is mainly dealing with
{\it nonlinear} negative friction, we restrict  ourselves
to the Gaussian thermostat (linear friction) and also discuss
white noise driving for completeness. The exact solution of the BGK model
with the full nonlinear friction is not known.

 In a crude scenario for relaxation without energy input, the
velocity distribution $F(v,t)$ relaxes towards a Maxwellian with shrinking
width $\alpha v_0(t)$, at a rate $\propto v^\nu_0(t)$. The width,
proportional to $\alpha$, mimics the role of the coefficient of
restitution, which reduces the typical velocity in an inelastic collision
by a factor $\alpha$. With a constant supply of energy, the system can
reach a NESS, and the global evolution can be modelled by the BGK-type
kinetic equation,

\begin{figure}
\includegraphics[angle=0,width=7cm]{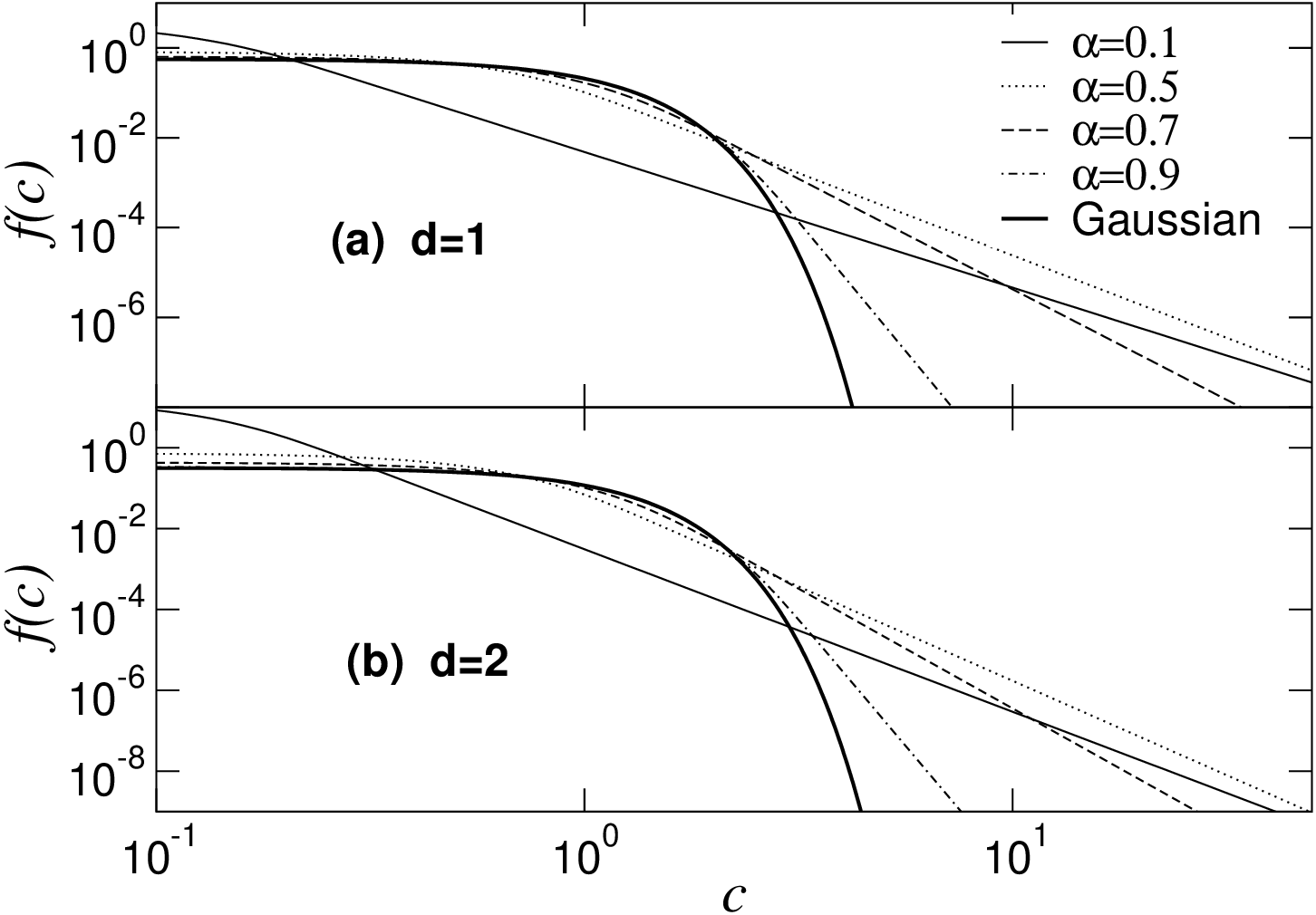}
\caption{BGK in one and two dimensions for FC: log-log plot
showing the power-law tails. } \label{bgkfc}
\end{figure}

\ba \Label{a1}
&&\partial_t F(v,t) -D\partial_{\bv}^2 F(v,t)=I(v|F)
\nn&& I(v|f) =
-v_0^{\nu}(t) [F(v,t)-F_0(v,t)] \nn &&F_0(v,t)= (\alpha v_0)^{-d}
\phi(v/\alpha v_0(t)),
\ea
where $\phi(c) =\pi^{-d/2} \exp[-c^2]$ is the Maxwellian. If $F(v,t)$ is
rapidly approaching the scaling form \Ref{scaling-f}, the rescaled
collision kernel, $I(v|F) = v_0^{\nu -d}I(c|f)$, takes the form,
\be \label{bgk-I}
I(c|f) = -f(c) + \alpha^{-d} \phi(c/\alpha) .
\ee
We note that the collision kernel does not show any $\nu$-dependence. This
is similar to Maxwell models, where the collision frequency is
independent of the {\it microscopic} velocities of the colliding
particles. The time evolution of $v_0(t)$ in free cooling and in the WN
(white noise) case obeys equations of motion, similar to
\Ref{v0}-\Ref{rate-v0-NF}. Because for inelastic soft spheres the rhs is
also proportional to $1-\alpha^2$, the discussion about stability of the
granular temperature $T(t) \sim v^2_0(t)$ is the same as in free cooling
and WN driving \cite{JSP-1,EPL}, and the same applies to Haff's
homogeneous cooling law, $ v^2_0 (t) \sim t^{-2/\nu}$.

In the free cooling case ($D=0$) the energy equation becomes,
$\dot{v}_0 =-2pq v_0^{\nu+1}$. Inserting
then the scaling ansatz \Ref{scaling-f} in \Ref{a1} yields,
\be\label{a4}
c f^\prime +(d+a) f =\frac{a}{\alpha^d} \phi\left(\frac{c}{\alpha}\right);
\qquad a = 2/(1-\alpha^2)
\ee
Its exact solution is (see also \cite{BMD,EB-SpringII}),
\ba\Label{a8}
f(c)&= & \frac{a \alpha^a}{\pi^{d/2}}
 \left(\frac{1}{c^{d+a}}\right)
\int_0^{\alpha c} du\, u^{d+a-1} e^{ -u^2} \nn &\sim& \frac{a
\alpha^a\Gamma\left( \frac{d+a}{2}\right)}{2\pi^{d/2}}
 \left(\frac{1}{c^{d+a}}\right) \quad (c\gg 1)
\ea
This solution, including its high energy tail (see Fig. \ref{bgkfc}), is independent of the
exponent $\nu$. A similar heavily overpopulated
power law tail, $f(c)\sim 1/c^{d+a}$ with $d>1$, is also found in the
freely cooling {\it Maxwell model}. There the exponent $a(\alpha)$ takes
in the elastic limit ($\alpha \to 1$) the very similar form $a \simeq 4d
/(1-\alpha^2)$ (compare \Ref{a4}). We also note that the
$\alpha$-dependence of the power law exponent in the BGK model is
essentially the same as for higher dimensional Maxwell models
\cite{JSP-1}, and similar to \Ref{S9} for NF driving. However, in the
general class of inelastic soft sphere models with collision frequency $
g^\nu$ and $\nu>0$ (hard scatterers),  the tail is not a heavily
overpopulated power law tail, but a lightly overpopulated stretched
exponential, $f(c) \sim\exp [-\beta c^b]$ with $b=\nu >0$. The BGK models
describe quite well the features of the soft scattering models, but are
totally missing the more effective randomization caused by the high speed
particles present, in models with positive $\nu$.

Let us now turn to the case of {\it white noise driving} in the BGK model
of Eqs. \Ref{a1}. Again the energy balance equation is the same as \Ref{v0}
for  soft spheres.  So all BGK models with $\nu >-2$ have a stable
attracting fixed point $v_0(\infty)$, and the integral equation
 has a rescaled form, analogous to \Ref{int}, \\
\be\Label{a10}
f''(c) +\frac{d-1}{c} f'(c) -2a f(c) = -\frac{2a}{\alpha^d} \phi\left(
\frac{c}{\alpha}\right).
\ee
Here a prime on $f$ denotes a derivative with respect to its argument $c$.
Eq. \Ref{a10} shows that $f(c)$ is independent of the model parameter
$\nu$ and of the noise strength $D$. The equation can be solved exactly,
and the two integration constants are
fixed by the normalizations \Ref{norm-f}. For all values of $d$ we make
the transformation
\be \Label{a11}
f(c) = \alpha^{-d} y(\beta c); \quad b = \half \beta \alpha; \quad \beta
=\sqrt{2a} =2/\sqrt{1-\alpha^2},
\ee
where $y$ is a function to be determined. The resulting equation 
for $y$ can be solved: 
the {\it one-dimensional} BGK model has the exact solution
\be \Label{BGK-1d}
y(x) = \half {b \exp[b^2]} \left[ e^x {\rm erfc}(b+\frac{x}{2b}) + e^{-x}
{\rm erfc}(b-\frac{x}{2b}) \right].
\ee
Using the properties of the complementary error function ${\rm erfc}(z)$
one can verify that the first term inside $[\cdots]$ decays for $ x \to
\pm \infty$ as $\exp[-x^2/4b^2]$ and the second one as $2\exp[-x]$,
yielding an exponential tail,
\be \Label{a12}
f(c) \sim  \half \beta \exp[b^2] e^{-\beta |c|}  \qquad (c\gg 1).
\ee

Similarly we find in the {\it two-dimensional} case for the solution
satisfying the normalizations \Ref{norm-f}, i.e.
\ba \Label{a13}
y(x) &=&\frac{1}{\pi} K_0(x) \int_0^x zdz \exp[-z^2/4b^2] I_0(z)
\nn
 &+& \frac{1}{\pi} I_0(x) \int_0^x zdz \exp[-z^2/4b^2]K_0(z),
\ea
where $I_0(x)$ and $K_0(x)$ are Bessel functions with imaginary argument
\cite{Abram+Stegun}. 
The exact solutions \Ref{a8}, \Ref{BGK-1d} and 
\Ref{a13} have been obtained by K. Shundyak
\footnote{Thanks are due to Kostya Shundyak for determining
the  exact solutions of the ODE's in this subsection using Mathematica.}.
At large $x$ we have $K_0(x) \sim e^{-x} \sqrt{\pi
/2x}$ and $I_0(x) \sim \exp[-x^2/4b^2]$, yielding the high energy tail,
\be \Label{a14}
f(c) \sim \frac{e^{\alpha^2/(1-\alpha^2)}}{\sqrt{\pi}(1-\alpha^2)^{3/2}}
\left(\frac{e^{-\beta c}}{\sqrt{c}}\right).
\ee
For higher dimensions $(d>2)$ we only quote the asymptotic
 solution,
\be \label{a18}
f(c) \sim  c^{1-d/2}e^{-\beta c} \ ,
\ee
which may also be obtained directly from \Ref{a10} by
neglecting the inhomogeneity, i.e. the gain term $ I_{\mbox{gain}}$ $\sim
\exp[-c^2/\alpha^2]$.

\begin{figure}
\caption{BGK with WN driving, in one and two dimensions, showing the
exponentially decreasing tails.
} \label{figbgkwn}
\end{figure}

Comparison with the results of \cite{JSP-1} for the WN-driven soft sphere
models shows that the large-$c$ behavior is exactly the same as that of
the Maxwell model ($\nu=0$), but the scaling solutions, displayed in Fig.
\ref{figbgkwn}, are independent of $\nu$ (since Eq. \Ref{a10} is itself
independent of $\nu$), whether the scaling solution is a stable attracting
state of a hard scattering model, or an unstable repelling state state of
a soft scattering model. It shows therefore again that the BGK model is
inadequate to model hard interactions.

In summary, the simple {\em linear} BGK model, although displaying
interesting features, such as power-law velocity distribution tails, is
far from being able to capture the rich behavior of the Boltzmann
equation, in particular it fails for hard interactions ($ \nu >0)$.

\section{Concluding remarks}
\label{sec:concl}

\begin{figure}
\includegraphics[angle=0,width=7cm]{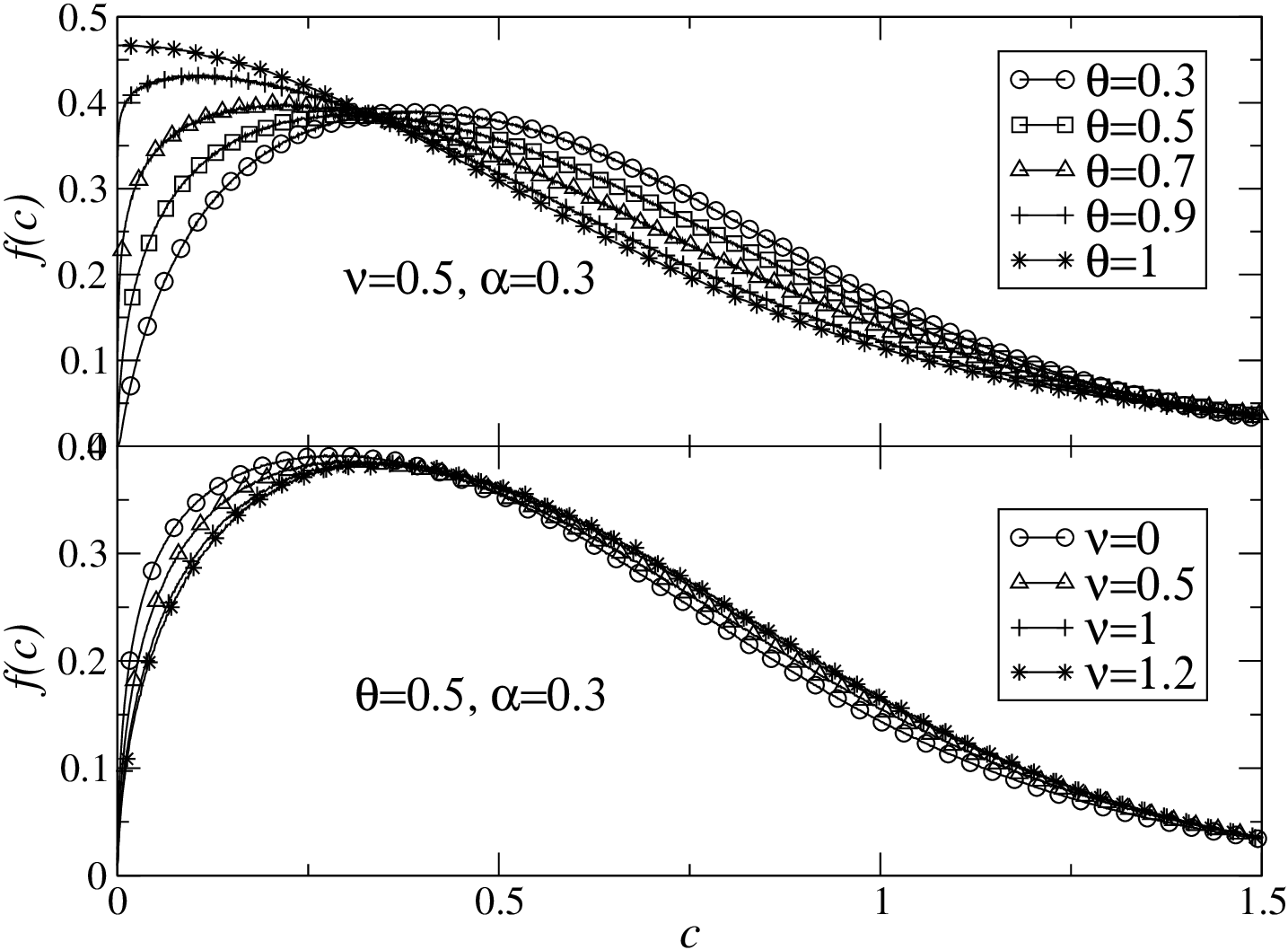}
\vskip .65cm
\includegraphics[angle=0,width=7cm]{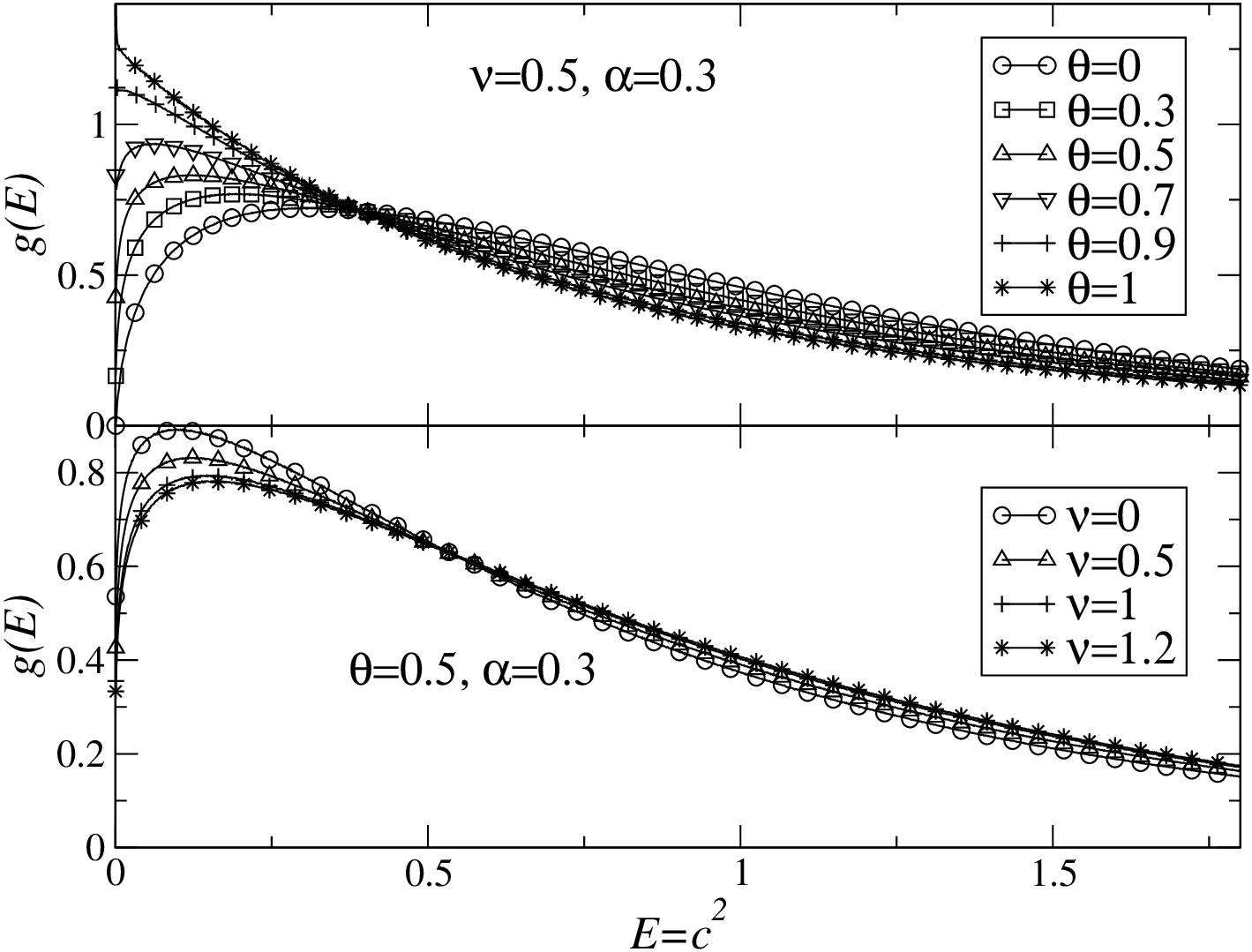}
\caption{Isobestic points as either $\theta$ is changed
at constant $\nu$ and $\alpha$, or as $\nu$ is changed
at constant $\theta$ and $\alpha$. Top: $d=1$; Bottom $d=2$
(in 2 dimensions we plot the distribution of the energy $E=c^2$).
} \label{figisobest}
\end{figure}
Within the framework of the nonlinear Boltzmann equation coupled to
stochastic or deterministic driving forces and 'heat' baths,
we have studied a general class of inelastic soft
sphere models. Our approach encompasses a broad class of
previously introduced models,
from hard scatterers like inelastic hard spheres (and even
very hard spheres \cite{Phys-Rep-ME}), to soft scatterers like Maxwell
molecules, and even softer ones with $\nu<0$, where $\nu$ governs
the dependence of collision frequency on relative velocity $g$
through a term $g^\nu$.

We have shown that the velocity
distribution $f(c)$ has a stretched exponential tail $\propto\exp(-c^b)$,
when the non-equilibrium
steady state is an attractive fixed point of the dynamics.
In certain regions of model parameters ($\nu,\alpha,\theta$)
where $\alpha$ denotes the restitution coefficient and $\theta$ is a friction
parameter, we have reported
important sub-leading corrections, where $f(c)$ is found to be of the form
$c^\chi\exp(-\beta v^b+\beta' v^{b'})$. The comparison with high-precision
numerical solutions of the Boltzmann equation, obtained through Monte Carlo
simulations (DSMC scheme), shows
that neglecting these sub-leading corrections in a fitting procedure can
lead to erroneous estimates of $\beta$. Algebraic distributions emerge in
cases of marginal stability ($b=0$), and we have calculated the corresponding
power law exponents. The high accuracy of our DSMC simulations have enabled us to
verify the theoretical predictions for a wide range of parameter values.

The models studied here are partially amenable to analytical progress,
but some features resist understanding. We conclude here by reporting
one such feature, that is illustrated in Fig. \ref{figisobest}.
We observe that all steady state rescaled
velocity distributions, at fixed $\nu$ and varying $\theta$,
pass through a common point. A similar property seems to hold
when $\theta$ is held fixed, and varying $\nu$. Such points, that can be coined
``isobestic'', have already been observed in a different context
(see e.g. section IV-E in reference \cite{PTD}), where their
occurrence could not be rationalized.

Acknowledgements We would like to thank a referee for valuable suggestions.

\end{document}